\documentclass[journal]{IEEEtran}
\usepackage{amsmath,amsfonts}
\usepackage{algorithmic}
\usepackage{algorithm}
\usepackage{tikz}
\usetikzlibrary{positioning,arrows.meta,fit}
\usepackage{array}
\usepackage{booktabs}
\usepackage[caption=false,font=normalsize,labelfont=sf,textfont=sf]{subfig}
\usepackage{textcomp}
\usepackage{stfloats}
\usepackage{url}
\usepackage{verbatim}
\usepackage{graphicx}
\usepackage{cite}
\usepackage{xcolor}
\usepackage{comment}
\usepackage{hyperref}
\usepackage[normalem]{ulem}
\usepackage{titlesec}
\titlespacing*{\subsubsection}{0pt}{0.5ex}{0.25ex}
\titlespacing*{\subsection}{0pt}{0.5ex}{0.25ex}
\titlespacing*{\section}{0pt}{0.5ex}{0.25ex}
\usepackage{etoolbox}
\makeatletter
\pretocmd\@bibitem{\color{black}\csname keycolor#1\endcsname}{}{\fail}
\newcommand\citecolor[1]{\@namedef{keycolor#1}{\color{blue}}}
\makeatother
\begin{document}
\bstctlcite{IEEEexample:BSTcontrol}

\title{A Robust 5G Terrestrial Positioning System with Sensor Fusion in GNSS-denied Scenarios}

\author{Hamed Talebian,
        Nazrul Mohamed Nazeer,
        Darius Chmieliauskas, 
        Jakub Nikonowicz,~\IEEEmembership{Senior~Member,~IEEE,}
        Mehdi Haghshenas,
        Łukasz Matuszewski,
        Mairo Leier, and
        Aamir Mahmood,~\IEEEmembership{Senior~Member,~IEEE}
\thanks{H. Talebian, N.M. Nazeer, and D. Chmieliauskas contributed equally.}
\thanks{H. Talebian, M. Haghshenas, and A. Mahmood are with the Department of Computer and Electrical Engineering, Mid Sweden Univerity, 851 70, Sundsvall, Sweden (email: firstname.lastname@miun.se)}
\thanks{N.M. Nazeer and M. Leier are with the Department of Computer Systems, Tallinn University of Technology, Estonia}
\thanks{D. Chmieliauskas is with the Department of Computer Science and Communication Technologies, Vilnius Gediminas Technical University, Lithuania.}
\thanks{J. Nikonowicz and Ł. Matuszewski are with the Faculty of Computing and Telecommunications, Poznan University of Technology, 60-995 Poznan, Poland (email: firstname.lastname@put.poznan.pl)}
\vspace{-20pt}
}



\maketitle

\begin{abstract}
This paper presents a terrestrial localization system based on 5G infrastructure as a viable alternative to GNSS, particularly in scenarios where GNSS signals are obstructed or unavailable. It discusses network planning aimed at enabling positioning as a primary service, in contrast to the traditional focus on communication services in terrestrial networks. Building on a network infrastructure optimized for positioning, the paper proposes a system that leverages carrier phase (CP) ranging in combination with trilateration to localize the user within the network when at least three base stations (BSs) provide line-of-sight (LOS) conditions. Achieving accurate CP-based positioning requires addressing three key challenges: integer ambiguity resolution, LOS/NLOS link identification, and localization under obstructed LOS conditions.
To this end, the system employs a multi-carrier CP approach, which eliminates the need for explicit integer ambiguity estimation. Additionally, a deep learning model is developed to identify NLOS links and exclude them from the trilateration process. 
In cases where LOS is obstructed and CP ranging becomes unreliable, the system incorporates an error-state extended Kalman filter to fuse complementary data from other sensors, such as inertial measurement units (IMUs) and cameras. This hybrid approach enables robust tracking of moving users across diverse channel conditions.
The performance of the proposed terrestrial positioning system is evaluated using the real-world KITTI dataset, featuring a moving vehicle in an urban environment. Simulation results show that the system can achieve a positioning error of less than 5 meters in the KITTI urban scenario—comparable to that of public commercial GNSS services—highlighting its potential as a resilient and accurate solution for GNSS-denied environments.
\end{abstract}

\begin{IEEEkeywords}
Carrier phase positioning,
LOS/NLOS identification, Error-state Extended Kalman Filter, visual odometry, 5G terrestrial positioning, sensor fusion.
\end{IEEEkeywords}

\section{Introduction}

Traditionally, the global navigation satellite system (GNSS) has been the primary technology used for positioning and navigation. It has reliably supported numerous applications for decades. However, recent events have exposed a critical vulnerability in GNSS-dependent systems. In particular, over large geographical areas, GNSS signals have been deliberately disrupted or denied through the emission of strong radio frequency noise, effectively rendering GNSS-based positioning inoperable~\cite{ghizzo2025assessing,borio2016impact,spanghero2025gnss}. Such distruptions highlight the urgent need for a more resilient and robust positioning system—one that does not rely exclusively on a single source of information. 
 
Cellular networks are increasingly being recognized as a promising infrastructure for providing localization services, offering a potential alternative/complementary to GNSS, especially in urban or indoor environments where GNSS signals may be obstructed or degraded. Historically, various positioning techniques have been explored within cellular systems to enable location estimation such as received signal strength (RSS), time of arrival (TOA), time difference of arrival (TDOA), angle of arrival (AOA), angle of departure (AOD), cell ID, and fingerprinting~\cite{chen2021carrier}. 

With the recent widespread deplpoyment of 5G networks several studies have been conducted to analyze the performance of different positioning schemes utilizing 5G networks~\cite{2019TDOAPSSSSS,ferre2019positioning,2017AOATOA,2020HenkAOATDOA,2022AOATOA,malmstrom20195g,2020Fingerprinting,gante2020deep}. For instance, in~\cite{2019TDOAPSSSSS}, primary synchronization signal (PSS) and secondary synchronization signal (SSS) in 5G is utilized to obtain TDOA, enabling user positioning. Similarly,~\cite{ferre2019positioning} focuses on estimating TDOA using dedicated positioning reference signals (PRS) to improve localization accuracy. In~\cite{2017AOATOA}, authors estimate TOA and AOA using uplink reference signal and fuse them together to achieve an accurate position estimate. In the same study, an Extended Kalman Filter (EKF) is employed to combine estimation from multiple BSs and improve localization performance. However, all these traditional geometric approaches rely on the assumption of a dominant line-of-sight (LOS) path — an assumption that may fail in real-world conditions. Their performance deteriorates significantly in rich multipath environments or under non-line-of-sight (NLOS) conditions~\cite{stephan2024angle}. To address these challenges, researchers in~\cite{malmstrom20195g,2020Fingerprinting,gante2020deep} have explored fingerprinting and machine learning (ML) techniques for positioning, particularly in scenarios where traditional methods struggle due to NLOS or dense multipath effects.


The growing importance of location-based services has prompted 3GPP to dedicate significant efforts toward developing and standardizing advanced positioning techniques for 5G and 5G-Advanced networks, particularly from Releases (Rel.)~16 through Rel.~19. These initiatives aim to enable centimeter-level positioning accuracy, which is critical for a wide range of emerging applications, including autonomous driving, industrial automation, and augmented reality~\cite{fouda,wang2023recent}. To support this goal, 3GPP specified two dedicated positioning signals in Rel.~16: the PRS for downlink-based positioning and the sounding reference signal (SRS) for uplink-based methods. Rel.~17 introduced multipath reporting, time delay correction between receiver and transmitter, and LOS/NLOS identification where the LOS/NLOS likelihood is utilized to weight certain measurements for localization~\cite{wang2023recent}. Rel.~18 expands upon these capabilities by introducing bandwidth aggregation to improve the effective bandwidth of the reference signal~\cite{Cha2025} and carrier phase positioning (CPP)—a highly precise technique that leverages the phase of radio frequency (RF) carrier waves to estimate the distance between user equipment (UE) and 5G base stations (BSs)~\cite{10644093}. By measuring the subtle phase shifts of these high frequency signals, CPP enables fine-grained distance estimation, making it a key enabler for achieving cm-level localization accuracy in next-generation networks.



\textcolor{black}{Building on established CPP principles and recent 5G-Advanced positioning developments \cite{nikonowicz2024,Abuyaghi2025,10644093,Cha2025,Saikko2025}, this paper presents an integrated terrestrial positioning system operating over existing 5G infrastructure. The system localizes the UE using CP measurements when at least three BSs provide LOS conditions and maintains continuity during LOS disruptions by fusing complementary sensors - inertial measurement units (IMUs) and camera-based visual odometry (VO). The paper focuses on system-level mechanisms required for reliable CPP in 5G deployments, including ambiguity handling under bounded-range conditions and link validity assessment in mixed LOS/NLOS environments.}

While our primary objective is to showcase the feasibility of achieving accurate user localization using 5G networks in practical scenarios, we also elaborate on the technical challenges encountered during system development and the solutions we implemented. These challenges and corresponding methodologies are summarized below:
\begin{itemize}
    \item \textbf{Carrier Phase Ranging:} \textcolor{black}{We leverage the high-precision nature of CP ranging for user localization in environments with multiple LOS-connected BSs. To resolve the critical challenge of integer ambiguity (IA)--the unknown number of full carrier wavelengths between the UE and BSs--we depart from methods that rely on external RAN assistance or hybrid data fusion~\cite{chen2021carrier,fan2021carrier}. Building upon established multi-frequency principles, we contribute a standalone multi-frequency framework that introduces: (i) an ambiguity-free distance configured via an adjustable virtual wavelength, and (ii) a noise correction mechanism based on phase progression analysis and enhanced phase averaging to improve measurement reliability. Consequently, this enables single-epoch, high-precision ranging inherently tailored to the operational environment.}
    \item \textbf{LOS/NLOS Classification Using ML:} Since CP ranging assumes LOS propagation, the presence of NLOS conditions can degrade accuracy. However, prior knowledge of the LOS condition needs to be obtained. To address, we develop a confidence-aware temporal convolutional neural network (T-CNN) module that analyzes channel impulse response (CIR) to detect-and-discard NLOS links in real time, enabling to selectively use only valid ranging data. \textcolor{black}{The main contribution in this area is data-driven downsampling with \textit{Stride} operation, multi-criteria model selection beyond plain accuracy score, and confidence region and value for soft decision.} 
    \item \textbf{Sensor Fusion with VO and IMU:} When LOS connectivity is lost, the system transitions to using auxiliary sensors. In particular, a camera-based visual odometry module is used to estimate motion and position, which is then fused with the last known CP-based location using EKF. Additionally, IMU data is integrated into the EKF to further enhance positioning robustness and accuracy during CP signal outages. This external information is utilized until the LOS condition is resumed. 
\end{itemize}

Above, we outlined our contribution to the development of an end-to-end positioning pipeline over 5G networks. The effectiveness of the proposed system is evaluated using the KITTI dataset~\cite{kitti}, which features a vehicle equipped with a camera, an IMU, and a GPS sensor navigating through urban streets under varying channel conditions, including LOS and NLOS scenarios relative to BSs. 
\textcolor{black}{Sionna ray tracing~\cite{hoydis2023sionna} is used to derive deployment-dependent LOS/NLOS visibility (and large-scale coverage trends) along the KITTI route; conditioned on these LOS/NLOS states, the PRS/SRS channel realizations used for CPP (and CIR inputs to the ML classifier) are generated using the 3GPP TR 38.901 UMi channel model}. Using the proposed methodology, we estimate the vehicle’s position and compare it against the standard GNSS-based positioning services. Under LOS conditions with at least three BSs, the CPP achieves cm-level accuracy. In contrast, when LOS is obstructed, the system transitions to relying on visual odometry and IMU sensors, which effectively maintain continuous position tracking. 
This hybrid approach validates the system's capability to provide seamless localization even in challenging NLOS environments.  Furthermore, we discuss the trade-offs, performance limitations, and practical considerations observed during these real-world evaluations, providing insights into the deployment feasibility and operational constraints of the proposed solution. 
\textcolor{black}{To ensure reproducibility, all scripts and experimental datasets are publicly available
\footnote{\url{https://github.com/aamirmahmood/SafeWork/tree/main}}.}
    
The rest of the paper is organized as follows. Section~\ref{sec:literature_reveiew} provides a comprehensive literature review, outlining the key problems and summarizing existing solutions. Section~\ref{sec:Architecture} establishes the architecture of the proposed positioning and tracking system. Section~\ref{sec:ran_aspects} offers an in-depth discussion on network planning and coverage, the characteristics of 5G channels, the structure of data and control signals.
Section~\ref{sec:cpp} introduces the theoretical foundations of CP ranging and details our proposed method for resolving the integer ambiguity. Section~\ref{sec:LOS_NLOS} then focuses on LOS/NLOS classification, presenting our ML-based identification framework. In Section~\ref{sec:sensor_fusion}, we describe the integration of visual odometry and IMU data through a sensor fusion approach based on an Error-State Extended Kalman Filter (ES-EKF) and evaluate the performance of the proposed end-to-end positioning system, including a comparison with GNSS-based positioning services. Finally, Section~\ref{sec:conclusion} concludes the paper by summarizing our key findings and outlining directions for future work.

\section{Relevant Works}
\label{sec:literature_reveiew}
The proposed positioning system comprises several key components, each contributing significantly to achieving a robust and highly accurate solution. This section breaks down these components and reviews relevant research associated with each. 

\subsection{Carrier Phase Positioning}
As highlighted earlier, CPP forms the foundation of the proposed system.
CPP has emerged as a pivotal technology for achieving cm-level accuracy in 5G-Advanced networks \textcolor{black}{\cite{nikonowicz2024,Saikko2025}}, driven by its inclusion in 3GPP Rel.~18 as a key enabler for mission-critical industrial IoT (IIoT) and smart city applications. 
The technical report \cite{3GPP_TR38_859} underscores CPP’s feasibility through system-level simulations, demonstrating cm-level accuracy under line-of-sight (LOS) conditions \cite{Cha2025}. However, practical implementations must deal with the ambiguity resolution challenge: the unknown integer number of full carrier cycles between UE and BSs. Recent works diverge in their approaches. For instance, Ou \textit{et al}. \cite{Ou2024} bypass direct ambiguity resolution by fusing TDOA with double-differenced CPDOA measurements. Their hybrid framework performs a localized search around TDOA-derived initial positions, mitigating clock offsets and phase wrapping effects through a weighted cost function. While computationally efficient, this method assumes reliable TDOA estimates—a limitation under severe multipath.  
In contrast, Deng \textit{et al}. \cite{Deng2024} tackle ambiguity resolution directly via double-differenced phase observations, isolating floating ambiguities using least-squares estimation and resolving integers through projection or Cholesky decomposition. Though rigorous, this approach introduces computational overhead, highlighting a key trade-off in CPP implementations. These challenges are also contextualized by Abuyaghi \textit{et al}. The survey~\cite{Abuyaghi2025} presents ambiguity resolution algorithms and emphasizes the need for robustness in non-ideal channel conditions—a concern echoed in \cite{3GPP_TR38_859}. 

However, despite these strides toward practical integration, the absence of holistic implementation frameworks persists—particularly for resolving ambiguities and validating performance in real-world deployments. Tools like the mixed-integer Cramér-Rao bound~\cite{Wymeersch2023} have emerged to bridge theoretical models and practical parameterization, yet they alone cannot address the procedural void in end-to-end CPP pipelines. Recent works sidestep systematic guidance for integrating disparate techniques, such as signal acquisition, ambiguity resolution, and error correction, into cohesive positioning systems. Collectively, these efforts highlight CPP’s maturation from theoretical promise to near-deployable capability. Nevertheless, the transition to ubiquitous cm-level positioning in 5G-Advanced networks hinges on unifying fragmented innovations into reproducible frameworks. Compared with recent advancements, \textit{our approach eliminates dependencies on external RAN measurements by resolving carrier-phase ambiguities entirely through multi-frequential phase analysis.}

\subsection{LOS/NLOS Classification}
\textcolor{black}{While CPP offers superior precision compared to traditional time-based localization methods, its performance degrades significantly under NLOS conditions. The accuracy gap between LOS and NLOS propagation is quantified in \cite{shah2025}, where UE positioning on DL-PRS and UL-SRS is compared in an urban deployment, showing that 
mean positioning error increases from $1.94\,\mathrm{m}$ (LOS) to $10.47\,\mathrm{m}$ (NLOS) for DL-PRS, and from $2.06\,\mathrm{m}$ (LOS) to $8.48\,\mathrm{m}$ (NLOS) for UL-SRS. It motivates measurement screening before localization via LOS/NLOS-aware measurement classification as a necessary ML-assisted step to maintain robust positioning performance, as specified by 3GPP \cite{3gpp_tr_38_857_v17_0_0}.}

\textcolor{black}{Several data-driven approaches are overviewded in \cite{pan2025ai}, including traditional ML, LSTM, and Transformer, while CNN is strong at local feature extraction and efficient for high-dimentional SRS input.\footnote{See \cite{Yu}, e.g., for non-data-driven LOS detection approach.} 
Additionally, CNNs are well-suited for CIR-based multipath channel classification because the CIR inherently exhibits recognizable local temporal structures in the delay domain. 
Unlike feature-engineering pipelines that compress the CIR into a small set of handcrafted statistics (e.g., power delay profile (PDP)-derived moments or threshold-based metrics), CNNs learn multi-scale filters that automatically detect such signatures directly from raw or lightly preprocessed CIR sequences. Moreover, CNNs are well-suited to CIR/CFR-based LOS/NLOS detection because their architectural priors—local receptive fields and weight sharing—reflect the structure of multipath propagation: discriminative signatures are often contained in local delay (or frequency) neighborhoods, and the same path-cluster patterns may occur at different delay indices and amplitudes across links. Thus, shared convolutional filters provide shift-tolerant feature extraction with low parameter complexity. This yields an accuracy--complexity advantage, since convolutional layers provide expressive feature extraction with parameter growth that is typically lower than fully connected architectures for high-dimensional inputs, enabling lightweight implementations suitable for real-time applications.}

\textcolor{black}{Consistent with standard CNN architectures, our proposed approach integrates a lower number of convolutional and fully connected layers, which reduces trainable parameters, utilizes learning-based feature summation via stride operations, and tunes the hyperparameters targeting the highest and lowest precision and iteration in addition to confidence-aware comparative inference analysis. A lightweight CIR-based CNN with a small classifier head for UWB NLOS/LOS identification, emphasizing deployment efficiency in \cite{si}, similar to our approach but with a relatively different architecture. An efficient lightweight SEL-CNN in \cite{zhu}, highlights the feasibility of low-complexity inference while maintaining competitive accuracy, although its model parameters are much higher than our approach, with marginal gain in accuracy. In \cite{wang}, one-dimensional wavelet packet analysis is integrated with a heavy 2D-CNN  to enable robust, precise channel identification, at the expense of 24\,h training time. A 1D-CNN is used for CIR denoising in \cite{Jiang} to boost LOS/NLOS classification performance, relatively similar to our proposal but with less accuracy. Building upon the established trend of adapting CNNs for multipath signal analysis, \textit{we contribute a high-performance fine-tuned architecture integrated as a system-level component to facilitate robust LOS/NLOS classification and filtering without altering the underlying appropriate CNN structure. Additionally, a confidence-aware $\kappa$ score beyond mere accuracy is adopted to evaluate the CNN prediction performance and select soft-exclusion of unreliable links before CPP}.}  

\subsection{Visual Odometry and Sensor Fusion for Positioning and Tracking}


As part of the proposed positioning framework, we integrate both visual and inertial sensors to enhance the positioning robustness and reliability, especially in scenarios where CPP inputs are unavailable. Here, we review recent works that consider fusing IMU and visual sensor data for positioning. 

Over the past decades, extensive research has been conducted on the integration of visual and inertial data for estimating pose—comprising position and orientation—a problem commonly referred to as visual-inertial odometry (VIO). VIO systems have been extensively applied in robotics, drones, and autonomous vehicles. State-of-the-art implementations such as~\cite{vins-mono,VIO-nonlinear-optimization,MSCKF-vio,scaramuzza2019visualinertialodometryaerialrobots} exemplify both filtering-based and optimization-based approaches. These systems typically rely on geometric constraints and known sensor models. Recently, learning-based VIO frameworks~\cite{ml-vio} have shown promise in improving estimation accuracy and robustness, especially under challenging visual conditions.

While these approaches deliver reliable short-term tracking, they are prone to cumulative drift in the absence of absolute references. To address this limitation, many methods incorporate additional sensors or map-based corrections to provide global constraints.
More recently, several studies have investigated the fusion of wireless signal measurements—such as Wi-Fi, Bluetooth, UWB, and cellular signals~\cite{uwb_vio}—as a means to mitigate long-term drift, offering an alternative to traditional map-based approaches for maintaining localization accuracy. However, such methods are often restricted to controlled indoor environments due to the scarcity of comprehensive real-world datasets that include wireless signal measurements.

Overall, sensor fusion has emerged as a central strategy for building resilient localization systems in environments where GNSS signals are degraded or absent. Combining multiple sensor modalities—including inertial measurements, visual features, and radio signals—significantly enhances accuracy, robustness, and continuity. 
\textit{To this end, we propose a positioning system that fuses visual odometry, inertial measurements, and CPP data using an ES-EKF.}

\section{System Design}
\label{sec:Architecture}
\begin{figure}
    \centering
    \includegraphics[width=1\columnwidth]{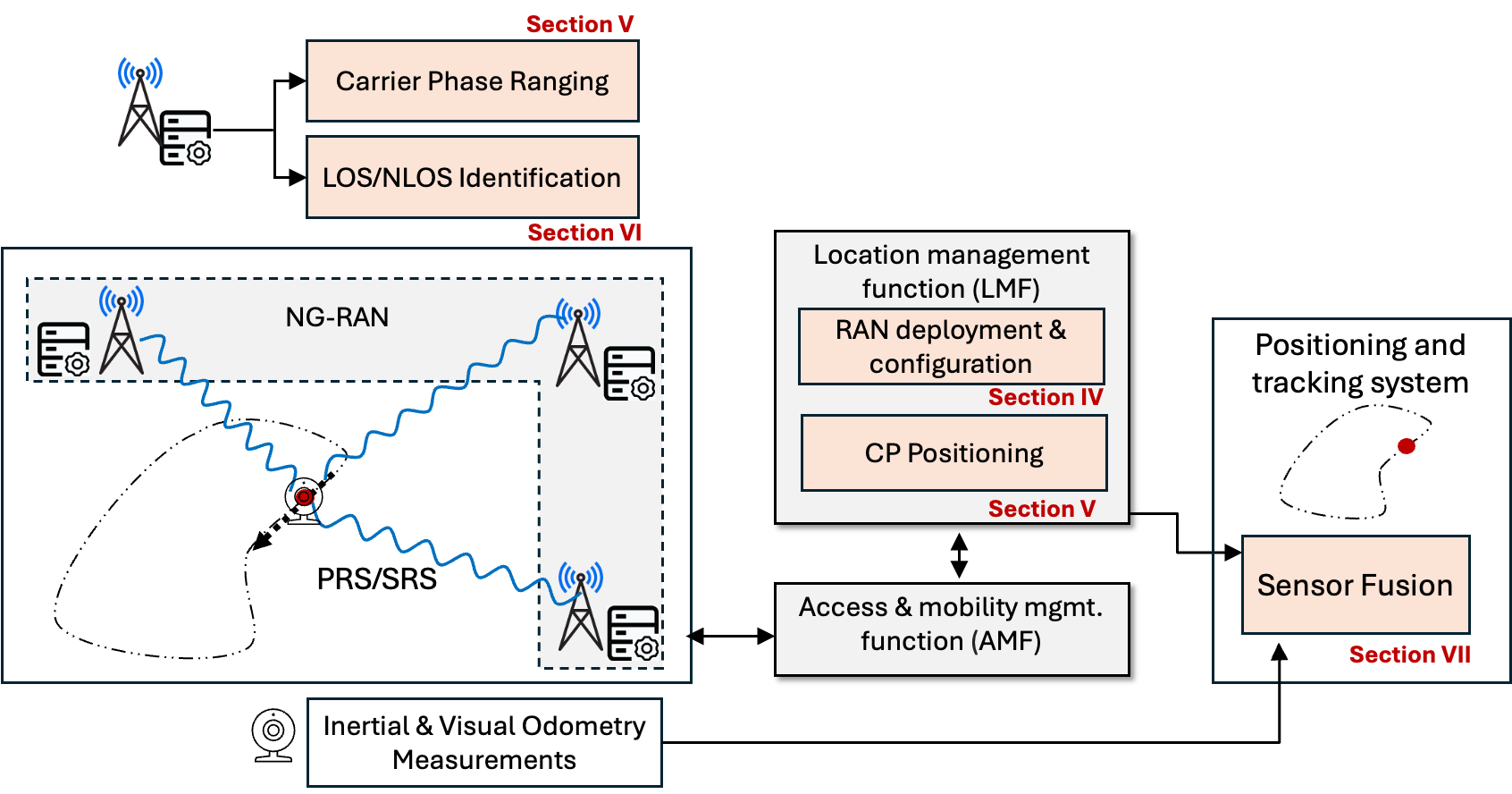}
    \caption{System-level depiction of the designed positioning and tracking system. The focused areas of the paper contributions, along with the relevant sections, are also shown.}
    \label{fig:SystemModel}
\end{figure}

This section presents a high-level overview of the proposed robust positioning framework, which integrates CPP in accordance with 3GPP standards, alongside a complementary sensor fusion module, as depicted in Fig.~\ref{fig:SystemModel}. While the system design aligns with the existing 3GPP positioning architecture, it is augmented with a fusion layer that incorporates visual odometry and IMU data to enhance reliability and continuity. This integration enables accurate and consistent position estimation, particularly in scenarios where LOS measurements are temporarily unavailable or degraded.

The 3GPP defines a modular positioning architecture consisting of the next-generation RAN (NG-RAN), the location management function (LMF), and the access and mobility management function (AMF), each with distinct responsibilities in enabling scalable and accurate localization. The NG-RAN comprises gNodeBs (gNBs), which transmit and receive dedicated positioning signals—PRS for downlink-based measurements and SRS for uplink-based measurements. The LMF, residing in the 5G Core (5GC), handles deployment-aware signal configuration and scheduling, collects measurement reports, and executes positioning algorithms. The AMF, also part of the 5GC, facilitates control-plane signaling, including configuring and coordinating positioning procedures between the LMF, NG-RAN, and UE.

The proposed design incorporates a CPP-based localization module that operates natively within the 3GPP framework using standard PRS and SRS signals. The system avoids the complexity of integer ambiguity resolution by using virtual wavelength processing, enabling efficient single-epoch position estimates. This approach ensures seamless integration with existing 5G NR protocols and infrastructure.

To support the precision of CPP in realistic propagation conditions, our system integrates a LOS/NLOS classification layer that filters measurements before positioning. Implemented as a lightweight deep learning model operating on SRS-derived CIRs, this component selectively admits only high-confidence LOS measurements. Its modular placement within the architecture allows for real-time classification of CP measurements.

Complementing the CPP engine, the sensor fusion module is designed to maintain positioning continuity during periods of degraded LOS visibility. By combining stereo visual odometry with IMU measurements in an ES-EKF, this module enables drift-resilient trajectory estimation. Its loosely coupled integration allows it to operate independently when needed while remaining synchronized with CPP when LOS signals become available again.


\textcolor{black}{The design uses Sionna ray tracing to evaluate deployment-dependent LOS/NLOS visibility and large-scale coverage along anticipated tracks for a selected gNB deployment configuration. It serves as the basis for determining the link-level PRS/SRS observations required for CPP (and CIR-based LOS/NLOS classification) using the 3GPP TR 38.901 UMi channel model, and for assessing the effectiveness of sensor fusion in LOS-deprived scenarios.}

By embedding our proposed CPP-based positioning method, LOS/NLOS classifier, sensor fusion capability, and deployment-aware management into the standard 3GPP architecture, we provide a robust, scalable, and centimeter-level accurate localization solution. This integrated approach ensures both the precision of CP-based positioning and the robustness required for practical, real-world deployments.

\section{RAN Deployment and Configuration for High-Precision Positioning}
\label{sec:ran_aspects}
All generations of RAN prior to 6G~\cite{giordani2020toward, wymeersch20226g} were designed primarily to support enhanced communication services, with localization considered a secondary feature~\cite{del2017survey}. Therefore, adapting RAN for high-precision, robust positioning requires specific adjustments to deployment and signaling configurations. 
These adjustments include deploying gNBs to improve LOS visibility during the RAN planning phase, and configuring physical layer (PHY) signals appropriately for high-resolution measurements. 

\subsection{Network Coverage Planning}
\label{sec:coverage}
Traditionally, a reliable network requires optimizing two key metrics during the planning phase: coverage and capacity. Effective radio network planning focuses on maximizing the signal-to-interference-plus-noise ratio (SINR) by minimizing inter-BS interference and eliminating coverage gaps. 
To achieve these goals, RF planning aims to minimize the overlap between the coverage footprints of adjacent BSs.

However, achieving high-precision positioning introduces additional requirements. Accurate positioning depends on the direct (i.e., LOS) reception of RF signals from multiple BSs~\cite{10644093}. Therefore, cellular network planning for positioning must prioritize BS placement to maximize LOS conditions. This dual optimization highlights the trade-off between minimizing interference for communication and ensuring sufficient LOS signal diversity for positioning. Alternatively, single-purpose PRS- or SRS-only transmission-reception points (TRPs), as described in~\cite{3gpp_ts_38_305_v18_3_0}, can be deployed.

\begin{figure}[ht]
\centerline{\includegraphics[width=\columnwidth]{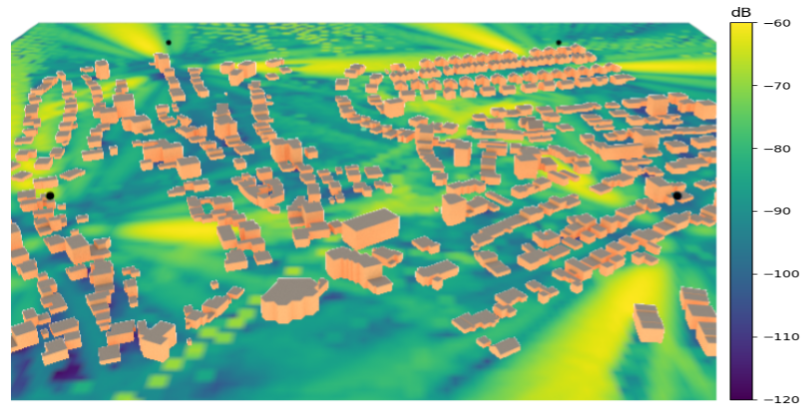}} 
\vspace{-10pt}
\caption{Simulated RF coverage in a suburban environment using four BSs. The map shows path gain levels in dB, calculated via ray-tracing including LOS, reflection, and diffraction effects.}
\label{fig:nlos_coverage}
\end{figure}

\begin{figure}[ht]
\centerline{\includegraphics[width=\columnwidth]{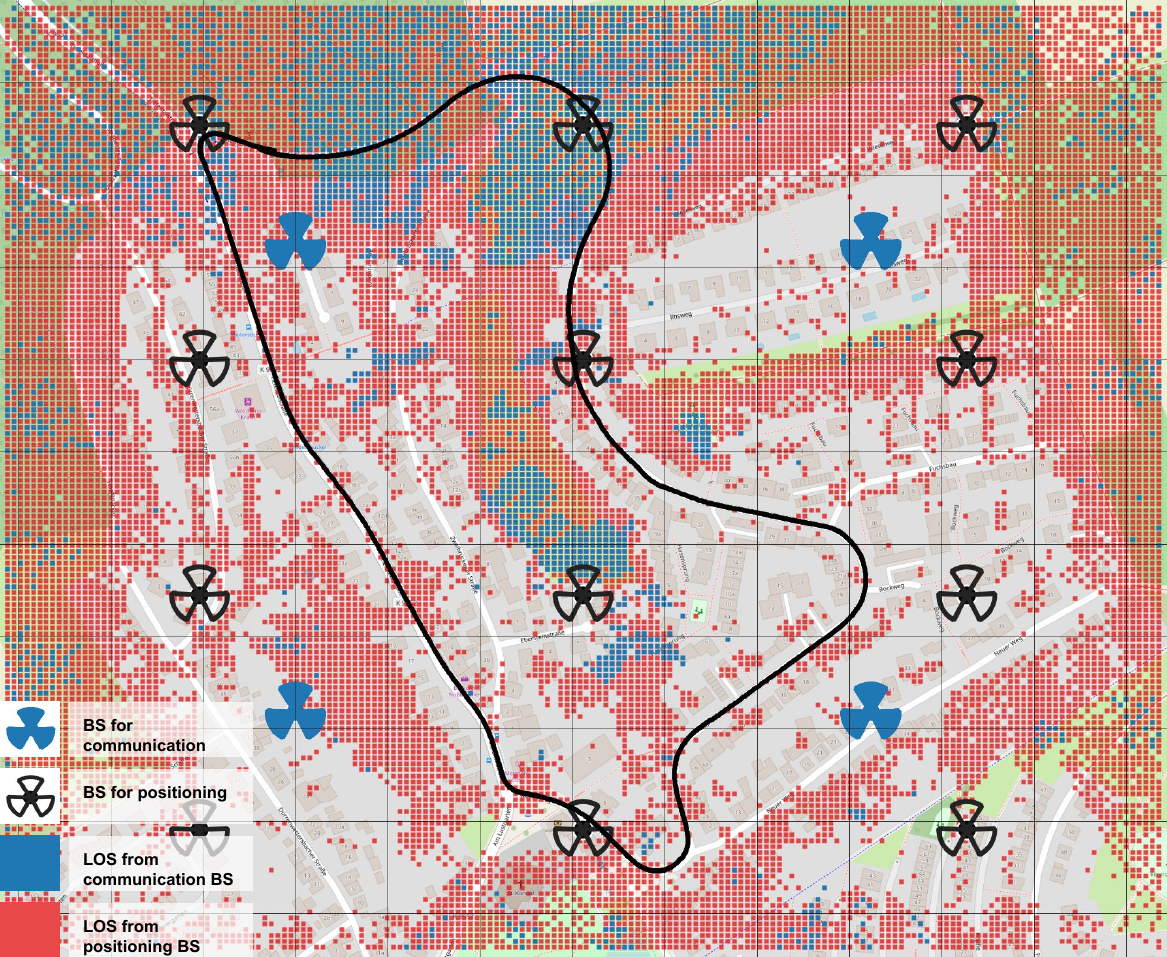}} 
\caption{Compares two network planning strategies: (1) The four blue cake-shaped symbols represent a conventional deployment designed primarily for communication services, resulting in limited LOS coverage—blue grid cells indicate regions with LOS to at least three BSs. (2) In contrast, the twelve black cake-shaped symbols correspond to a network layout optimized for localization, where positioning is treated as a primary service. }
\label{fig:los_3}
\end{figure}

To illustrate the contrasting requirements of RF coverage planning for communication and positioning, we performed ray-tracing simulations using Sionna. The resulting coverage levels for communication and positioning are shown in Fig.~\ref{fig:nlos_coverage} and Fig.~\ref{fig:los_3}, respectively.

Fig.~\ref{fig:nlos_coverage} shows a simulated RF coverage map for the suburban residential area in which the KITTI dataset~\cite{kitti} was collected, covering an area of $900\,\text{m}\!\times\!800\,\text{m}$.
Four BSs were positioned at a height of 30\,m—above rooftop level—and tilted downward by $6^{\circ}$. The simulation was performed using 3D building models and a ray-tracing engine that accounts for LOS, reflection, and diffraction effects. 
The resulting coverage map displays path-gain values on a $5\,\text{m}\!\times\!5\,\text{m}$ grid. As shown, the four BSs provide sufficient coverage across the entire area.


In contrast, Fig.~\ref{fig:los_3} illustrates the LOS visibility requirement for high-accuracy positioning, where a minimum of three BSs are in LOS. The simulation considered only direct LOS propagation, excluding reflection and diffraction effects. Following the simulation, each $5\,\text{m}\!\times\!5\,\text{m}$ grid cell was analyzed to determine how many BSs had a direct LOS path to that location. Blue grid cells indicate areas where at least three out of four deployed BSs simultaneously provides LOS coverage. In this scenario, only $16\%$ of the total area meets this requirement. These results emphasize the need to reconsider RF planning strategies when accurate positioning is a primary design goal.

To improve positioning coverage, we propose a denser deployment strategy tailored specifically for localization rather than conventional communication requirements. By increasing the number of BSs from four to twelve (depicted by black markers), LOS coverage to at least three BSs rises to approximately $64\%$ of the area. This enhanced layout meets the typical conditions required for accurate carrier-phase positioning, thereby demonstrating the importance of RF planning with localization objectives in mind.



When the objective is to enable high-precision positioning, simulation tools should be employed to determine the optimal placement of BSs, maximizing LOS probability throughout the coverage area. 
Despite the increased BS density, the simulation reveals several zones that still lack LOS visibility to three or more BSs. This highlights the importance of preprocessing algorithms capable of distinguishing LOS from NLOS paths, as well as the need to integrate sensor fusion techniques to enhance the robustness and accuracy of positioning solutions.


\subsection{5G NR PHY Channels and Signals suitable for Positioning}
\label{sec:channels_and_signals}
A fundamental requirement for any positioning system is prior knowledge of the signals used for localization.
In the NR PHY structure, several channels and signals with deterministic properties are identified. 
These include the NR common signals, synchronization signal block, and channel state information reference signal, which are used primarily for communication, as well as positioning-specific signals.







To support improved positioning accuracy, 3GPP has specified two dedicated signals: PRS and SRS. PRS is a downlink, single-purpose positioning signal, while SRS is primarily used for uplink channel measurements. Starting with 3GPP Rel.~16, the SRS is explicitly engineered to support positioning in addition to its original channel-sounding role.

\subsubsection{PRS Characteristics-- Periodicity,  Muting, and  Sequence Generation}

PRS is designed to overcome limitations in common reference signals, such as interference and poor correlation properties. This purpose-specific design improves the positioning, at the expense of introducing downlink communication overhead. 
PRS characteristics are highly configurable in time, bandwidth, and comb pattern allocation within the Physical Resource Block (PRB) as described in~\cite{3gpp_ts_38_214_v18_4_0}.

The network can tune PRS by adjusting the configuration parameters:
(i) the transmission period (4–10240 slots);
(ii) the number of consecutive OFDM symbols per slot;
(iii) the occupied bandwidth (24–272 PRBs, in 4-PRB steps);
(iv) the comb size, i.e., the number of subcarriers in each PRB that carry PRS;
(v) a binary muting pattern of length 2–32 that improves SINR by coordinating gaps among cells; and
(vi) one of 4096 Gold-sequence IDs~\cite{3gpp_ts_38_211_v18_4_0}.









PRS is specifically configured to ensure high Resource Elements (REs) density, improving positioning accuracy and detectability. However, the additional overhead caused by data transmission and the requirement for coordination among multiple BSs makes PRS optional/rare in commercial communication-oriented  networks.

\subsubsection{SRS Characteristics-- Periodicity, Scheduling, and Sequence Generation}

Similar to the PRS, 3GPP introduced SRS as an uplink signal for channel quality estimation and positioning purposes. 
SRS is configured individually for each UE to support channel estimation for communication. However, when used for positioning, it supports customized, positioning-specific configurations.

Key SRS parameters for positioning include~\cite{3gpp_ts_38_214_v18_4_0,3gpp_ts_38_211_v18_4_0,3gpp_ts_38_305_v18_3_0}: (i) periodicity and timing, where operators choose the number of OFDM symbols and triggering mode (periodic, semi-persistent, or on-demand) to balance uplink load against update rate; (ii) bandwidth, selected from the resource sets listed in \cite[Table~6.4.1.4.3-1]{3gpp_ts_38_211_v18_4_0}; (iii) comb spacing, with subcarrier densities of 2, 4, or 8; (iv) tight scheduling so that multiple BSs receive the UE’s SRS in the same time-frequency resource; and (v) sequence IDs, drawn from up to 65535 Zadoff–Chu sequences for orthogonality.

\begin{figure}[ht]
\centerline{\includegraphics[width=0.8\columnwidth]{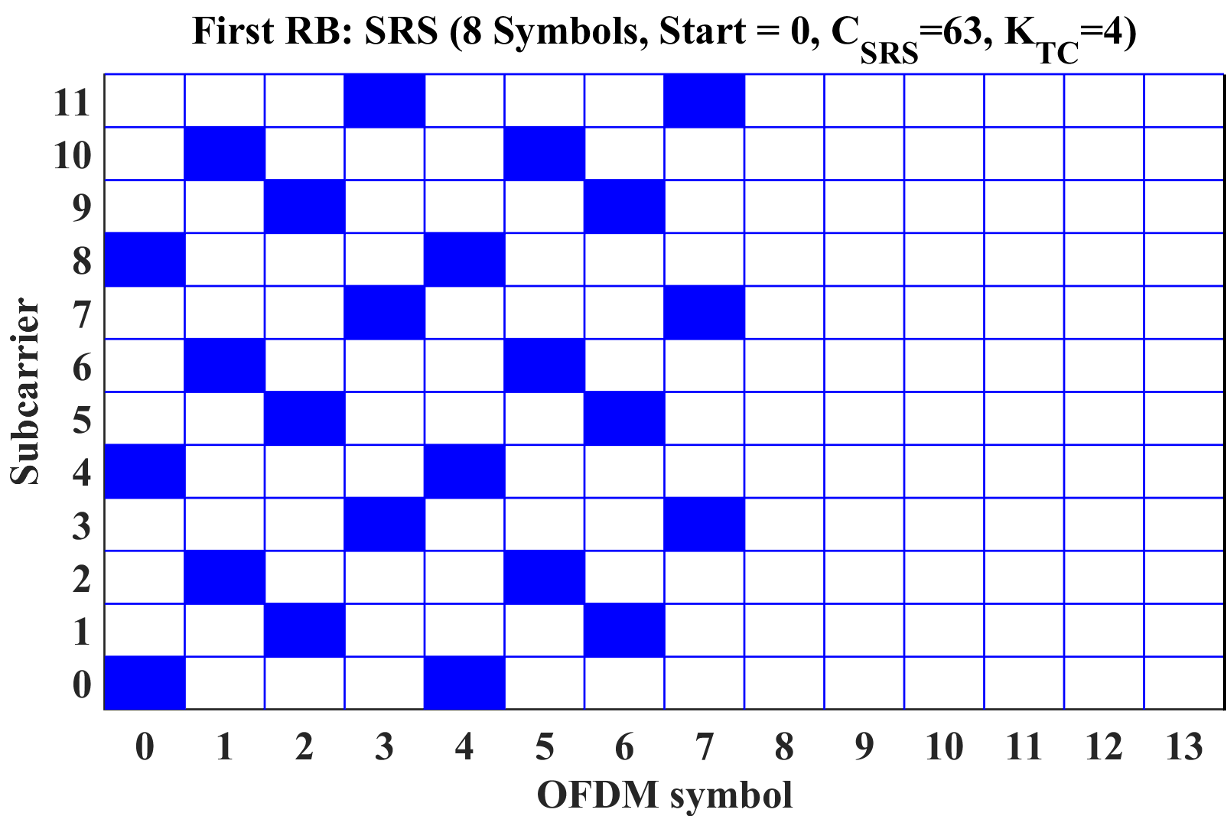}} 
\vspace{-8pt}
\caption{Example SRS configuration within the NR OFDMA grid. SRS allocation within a single PRB, illustrating comb-based subcarrier mapping across OFDM symbols.}
\label{fig:srs_ofdma_grid}
\end{figure}

Fig.~\ref{fig:srs_ofdma_grid} illustrates an example of SRS configurations within the NR OFDMA time-frequency grid.
It shows the SRS allocation within a single PRB, highlighting the occupied OFDM symbols and subcarriers based on a comb-type structure. 
SRS resources are allocated to a single UE by the gNodeB scheduler. Other UEs within the coverage area can be allocated similar SRS resources by applying different time and frequency shifts.

By configuring SRS with positioning-centric parameters (e.g., higher bandwidth or denser comb spacing), networks can enhance the accuracy of UL-based positioning methods. However, similar to PRS, these uplink-specific SRS configurations require additional uplink resources and coordination among BSs, balancing the trade-off between communication throughput and high-precision localization.

Both PRS and SRS can be generated using stable reference sequences and tightly controlled timing, making them well-suited for CP-based positioning. Their deterministic properties support high-fidelity phase measurements, enabling the subwavelength accuracy required for advanced, cm-level positioning.





\section{Carrier Phase Positioning in Challenging Propagation Environments}
\label{sec:cpp}

We propose a CP-based positioning framework for 5G NR systems, designed to meet the stringent accuracy requirements of industrial and commercial applications across both FR1: 0.41--7.125\,GHz and FR2: 24.25--52.6\,GHz, as specified in 3GPP Rel.~18 \cite{3GPP_TR21_918}. While initially targeting InF deployments, the methodology addresses general harsh propagation environments, including urban microcells (UMi), indoor offices (InO), and other scenarios characterized by dense clutter, NLOS, and dynamic obstacle configurations per 3GPP TR~38.901 channel models \cite{3GPP_TR38_901}.

The proposed CPP approach shares fundamental technical challenges with GNSS solutions \cite{Li2021_GNSSRepeater}, particularly IA resolution and multipath-induced phase errors. However, NR systems provide distinct advantages through controlled BSs deployments and multi-subcarrier capabilities that are unavailable in satellite navigation. Additionally, NR systems offer improved error mitigation techniques such as LOS/NLOS identification, enhancing the reliability and robustness of CP measurements. These capabilities support the application of CPP both as a standalone solution and as an enhancement to legacy NR positioning methods based on timing, angle, and power measurements, significantly improving overall accuracy.

\begin{figure}[t]
    \centering
    \includegraphics[trim={0 0 0 0.8cm},clip, width=0.5\textwidth]{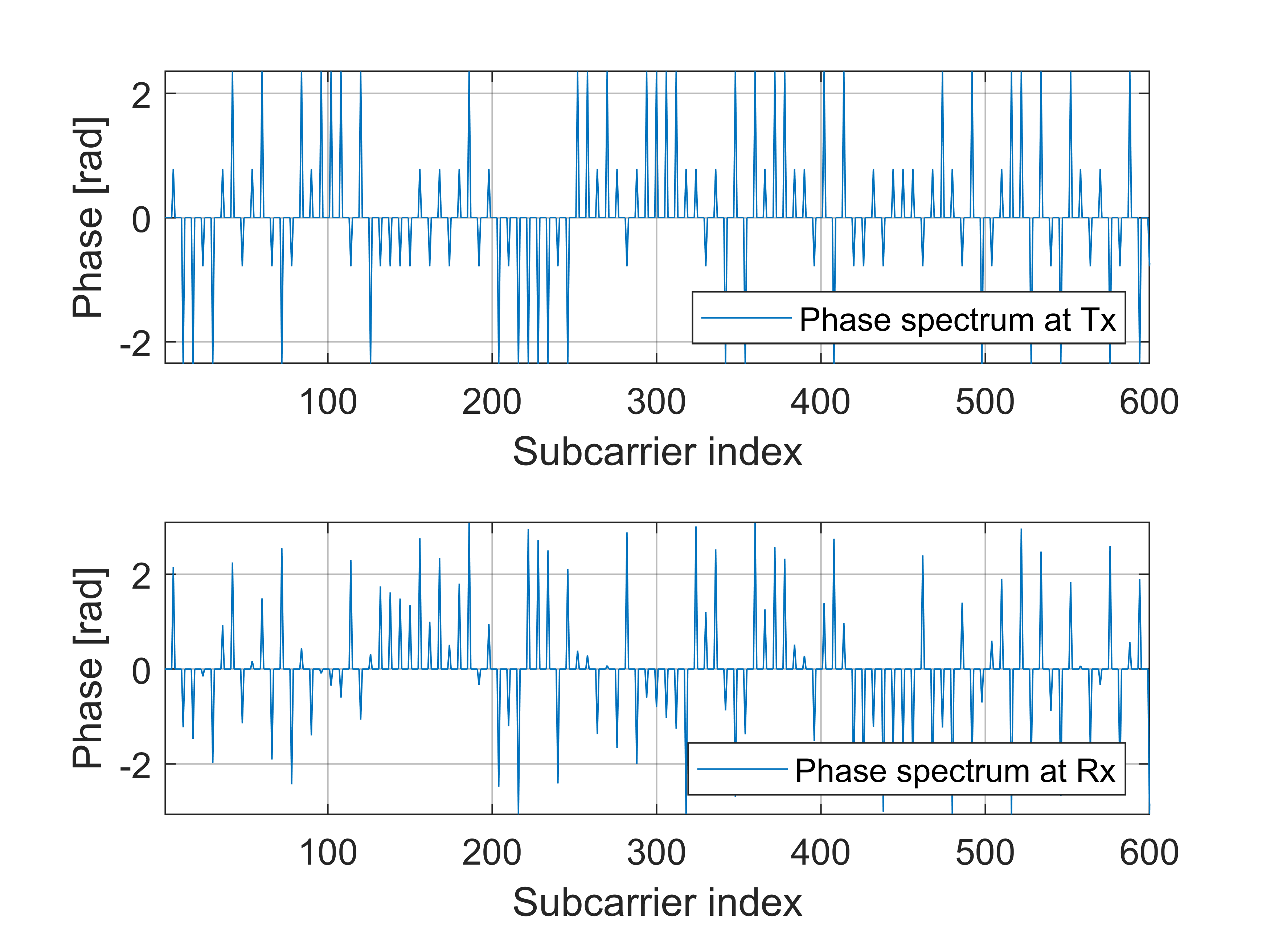}
    \vspace{-20pt}
    \caption{Phase spectrum at the transmitter and receiver, showing the phase shifts resulting from the propagation and communication channel.}
    \label{fig:raw_spectrum}
\end{figure}

CP measurements fundamentally determine the phase shift between an incoming carrier signal and a receiver's predefined reference phase (Fig.~\ref{fig:raw_spectrum}). The measured phase shift is confined to $[0, 2\pi)$ radians (or equivalently mapped from $[-\pi, \pi)$), with errors typically limited to a small fraction of the signal wavelength \cite{3GPP_R1-1901980}. Under LOS conditions, this enables precise distance calculation through linear phase-to-distance conversion, allowing cm-level positioning accuracy \cite{nikonowicz2024}. However, three critical limitations arise: (1) strict synchronization requirements between transmitter and receiver for phase comparison validity, (2) dependence on LOS/semi-LOS propagation to avoid multipath-induced distortions and timing ambiguities (NLOS scenarios degrade precision by orders of magnitude), and (3) the fundamental IA problem--the inability to resolve the integer number of full wavelength cycles ($N$) preceding the measured fractional phase. These constraints collectively challenge CPP's reliability in complex environments, despite its theoretical advantages for industrial/urban applications.
\subsection{Virtual wavelength}
\label{sec:virtual_wavelength}
{\color{black}
Let us consider a scenario in which a transmitter (Tx) emits two pilot symbols $x_1, x_2 \in \mathbb{C}$ over two distinct carrier frequencies $f_1$ and $f_2$. The corresponding transmitted passband signals can be expressed as
\begin{align}
\begin{split}
    & X_1 = x_1\cdot e^{-j( 2\pi f_1 t + \theta_{\mathrm{Tx}})} \\
    & X_2 = x_2 \cdot e^{-j( 2\pi f_2 t + \theta_{\mathrm{Tx}})} 
\end{split}
\end{align}
where $\theta_{\mathrm{Tx}}$ denotes the phase offset of the transmitter's local oscillator, which is assumed to remain constant across frequencies.
Assuming a LoS propagation environment, the corresponding received baseband signals at the receiver (Rx) can be modeled as
\begin{align}
\begin{split}
\label{eq:received_symb}
    & y_1 = \alpha_1 \cdot x_1 \cdot e^{-j( 2\pi f_1 \tau + \theta) } \\
    & y_2 = \alpha_2 \cdot x_2 \cdot e^{-j( 2\pi f_2 \tau + \theta) } 
\end{split}
\end{align}
where $\theta = \theta_{\mathrm{Tx}} - \theta_{\mathrm{Rx}}$ represents the phase offset between the transmitter and receiver local oscillators, assumed frequency-independent. The parameters $\alpha_1, \alpha_2 \in \mathbb{R}$ denote channel attenuation coefficients, and $\tau = \frac{d}{c}$ is the propagation delay, with $d$ being the Tx–Rx distance and $c$ the speed of light.

The received signals in \eqref{eq:received_symb} contain both the transmitted pilot symbols and the channel-induced phase rotation. To enable carrier-phase-based ranging, the contribution of the pilot symbols is removed by multiplying each received signal by the conjugate of the corresponding transmitted symbol, e.g., $\frac{x_1^* y_1}{|x_1|}$. This operation isolates the channel-induced phase component. The resulting phase, wrapped within the interval $(-\pi,\pi]$ contains information about the propagation delay and can be expressed as
\begin{align}
\begin{split}
    \phi_1 + 2\pi N_1 = 2\pi f_1 \frac{d}{c} + \theta = 
    \frac{2\pi d}{\lambda_1} + \theta \\
     \phi_2 + 2\pi N_2 = 2\pi f_2 \frac{d}{c} + \theta = 
    \frac{2\pi d}{\lambda_2} + \theta
\end{split}
\end{align}
where $\lambda_1 = \frac{c}{f_1}$ and $\lambda_2 = \frac{c}{f_2}$ and $N_1,N_2 \in \mathbb{Z}$ represent integer ambiguities arising from phase wrapping. 
To eliminate the unknown oscillator phase offset $\theta$, the phase difference between the two observations is computed as $\Delta\phi = \phi_{2}-\phi_{1}$, yielding
\begin{align}
\label{eq:phase_shift}
    \Delta \phi + 2\pi \Delta N = 2\pi d \cdot \frac{\Delta f}{c}= 2\pi d \cdot (\frac{1}{\lambda_2} - \frac{1}{\lambda_1}) = \frac{2\pi d}{\lambda_v}
\end{align}
where $\Delta f = f_2 - f_1$, $\Delta N = N_2 - N_1$, and the virtual wavelength $\lambda_v$ is defined as
\begin{equation}
    \label{eq_lambda_new}
    \lambda_{v} = \frac{1}{\frac{1}{\lambda_{2}}-\frac{1}{\lambda_{1}}} = \frac{c}{\Delta f}.
\end{equation}
Eq.~\eqref{eq:phase_shift} can be interpreted as the phase shift of an equivalent wave with wavelength $\lambda_v$ propagating between the transmitter and receiver. Accordingly, the propagation distance can be estimated as
\begin{align}
\label{eq:distance_estimate}
    d = \left(\frac{\Delta \phi}{2\pi} + \Delta N\right)\lambda_v
\end{align}
where $\Delta N$ denotes the IA that must be resolved within a finite set of candidates.

The concept of virtual wavelength provides an important advantage: by selecting an appropriate frequency spacing $\Delta f$, the virtual wavelength can be significantly larger than the individual carrier wavelengths (c.f., Fig.~\ref{fig:virtual_wave}). This expansion increases the unambiguous range, potentially allowing the true distance $d$ to lie within a single wavelength of the virtual wave, thereby eliminating the need for explicit IA resolution.

However, this benefit introduces a trade-off. As $\lambda_v$ increases, the distance estimate becomes more sensitive to phase measurement errors, such that small phase deviations result in larger ranging errors. Therefore, $\Delta f$ must be carefully chosen to balance ambiguity resolution and estimation accuracy.

As a practical example, consider a 5G system with a bandwidth of $400\,$MHz and subcarrier spacing (SCS) $\Delta f_{\mathrm{scs}} = 120\,$ kHz. In a comb-6 SRS configuration, pilots are transmitted on every sixth subcarrier, resulting in a frequency spacing of $\Delta f = 6 \cdot \Delta f_{\mathrm{scs}} = 720\,$ kHz. This corresponds to a virtual wavelength of approximately $416\,$m. Consequently, as long as the distance between the UE and BS is less than $\lambda_v$, the phase-difference measurement remains inherently ambiguity-free. Since typical InF/UMi link distances fall within this bound (considering the typical Inter-Site Distance of 200~m for UMi~\cite{3GPP_TR38_901}), explicit IA resolution is not required under such deployment conditions. 

\begin{figure}[h]
    \centering
    \includegraphics[width=0.5\textwidth]{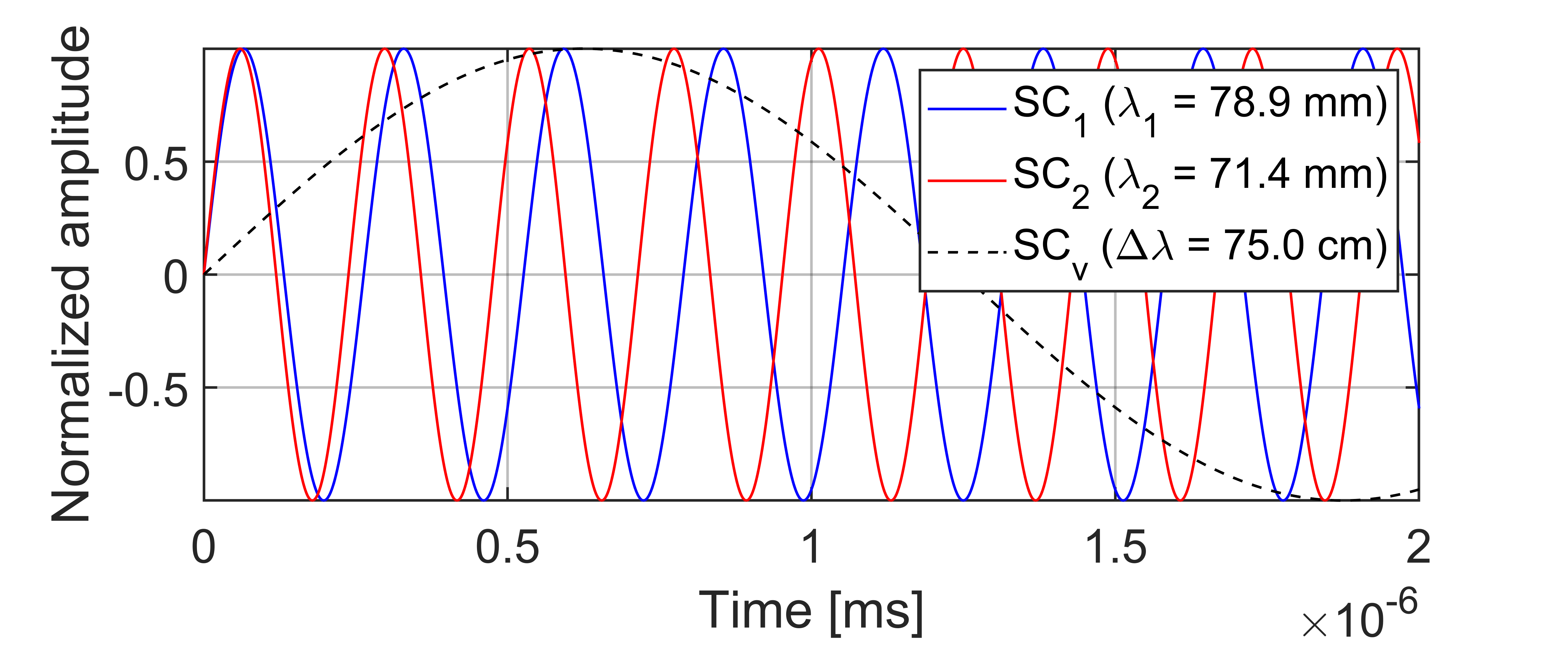}  
    \caption{Virtual wave from phase difference $\Delta\Phi$ of two subcarriers $\mathrm{SC}_{1}$ and $\mathrm{SC}_{2}.$}
    \label{fig:virtual_wave}
\end{figure}

\subsection{Solution outline}

The preceding discussion considered carrier-phase-based ranging using differential phase measurements between two frequencies. To improve robustness, mitigate noise effects, and reduce sensitivity to small-scale channel variations, it is advantageous to exploit observations across multiple subcarriers.

Considering a comb-type subcarrier allocation, as commonly employed in PRS or SRS configurations, let us assume that $K$ subcarriers are available. To perform differential phase analysis, a fixed spacing of $k$ subcarriers is introduced, where $0 < k < K/2$. By forming pairwise combinations across the subcarriers, a total of $K-k$ phase differences can be obtained as $\Delta\phi_{i} = \phi_{i+k} - \phi_{i}$, for $i = 1, \dots, K-k$.

To aggregate these measurements, \cite{R1-2306873} proposes estimating the average phase difference via the argument of a vector sum of complex subcarrier coefficients, effectively an amplitude-weighted circular mean. Since phase information is inherently encoded in the complex representation of each subcarrier, this method is computationally efficient and readily implementable. The corresponding formulation is given by
\begin{equation}
    \Delta\phi_{\text{avg}} = \text{arg} \left( \sum_{i=1}^{K-k} A_{i+k}e^{j\phi_{i+k}} -  A_{i}e^{j\phi_{i}}\right),
    \label{old_delta}
\end{equation}
where $\Delta\phi_{\text{avg}}$ is the average phase difference and $A_{i}e^{j\phi_{i}}$ denotes the complex representation of the 
$i$-th subcarrier. 

While this approach is statistically robust in environments dominated by dense, diffuse scattering, where deep fades with negligible amplitude naturally marginalize random phase noise, it proves suboptimal in InF or UMi scenarios as defined in the 3GPP channel models \cite{3GPP_TR38_901}. In these environments, characterized by a sparse multipath structure with a limited number of discrete clusters and high-energy specular components , the low number of effective scatterers (specifically $25$ for InF and $12$ for UMi LOS) results in phase contributions that lack the statistical dispersion required for mutual cancellation in the complex plane. Consequently, a single dominant reflection can persist across a broad range of subcarriers with high amplitude but a biased phase, systematically biasing the resultant vector in \eqref{old_delta} relative to the true LoS phase. Ultimately, the method ignores the strict geometric structure of phase differences by treating them as independent statistical variables. This lack of explicit alignment with the physical model in Section~\ref{sec:virtual_wavelength} makes the estimate suboptimal from a model-consistent perspective.

To address these limitations, we propose an alternative approach based on linear averaging of the extracted phase differences. By processing the phases directly, the estimate becomes invariant to the power-weighted biases typical of deterministic reflections. The proposed scheme is defined as:
\begin{equation}
\begin{aligned}
    \Delta\phi_{\text{avg}} = \frac{1}{K-k}\Bigl( \sum_{i=1}^{K-k} \text{arg}\left(A_{i+k}e^{j\phi_{i+k}}\right) -  \text{arg}\left(A_{i}e^{j\phi_{i}}\right) \Bigr).
    \label{new_delta}
\end{aligned}
\end{equation}
A critical aspect of implementing \eqref{new_delta} is the robust handling of phase differences, which inherently fall within the $(-2\pi, 2\pi]$ interval. Since measurements are wrapped within $(-\pi, \pi]$, direct subtraction leads to ambiguities that fail to preserve the direction of phase evolution across frequency. To ensure consistency with the underlying propagation physics, we enforce a monotonic phase progression constraint: the phase of a higher-frequency subcarrier must be advanced relative to the preceding one, reflecting the physical delay $\tau$ where $\phi(f) = -2\pi f \tau$. 
This procedure corresponds to phase unwrapping. Consequently, the phases in \eqref{new_delta} refer to unwrapped values.
In this approach, by tracking the local phase slope against the expected deterministic trend and enforcing the phase progression constraint, specular components and sparse fading effects causing significant local phase shifts can be identified and rejected as outliers. This mechanism allows the coherent LoS group delay to be isolated from inconsistent samples, while low-amplitude additive noise that introduces only small-variance distortions can be effectively average out across the remaining subcarrier set. Consequently, the final estimate is no longer skewed by high-energy reflected components inherent in InF or UMi environments. By incorporating phase progression condition, the proposed method maintains consistency with the physical model illustrated in Fig.~\ref{fig:oc_nz_pp_spectrum}, thereby improving the reliability and interpretability of the estimated phase differences.

\begin{figure}[h]
    \centering
    \includegraphics[trim={0 0 0 0.8cm},clip, width=0.5\textwidth]{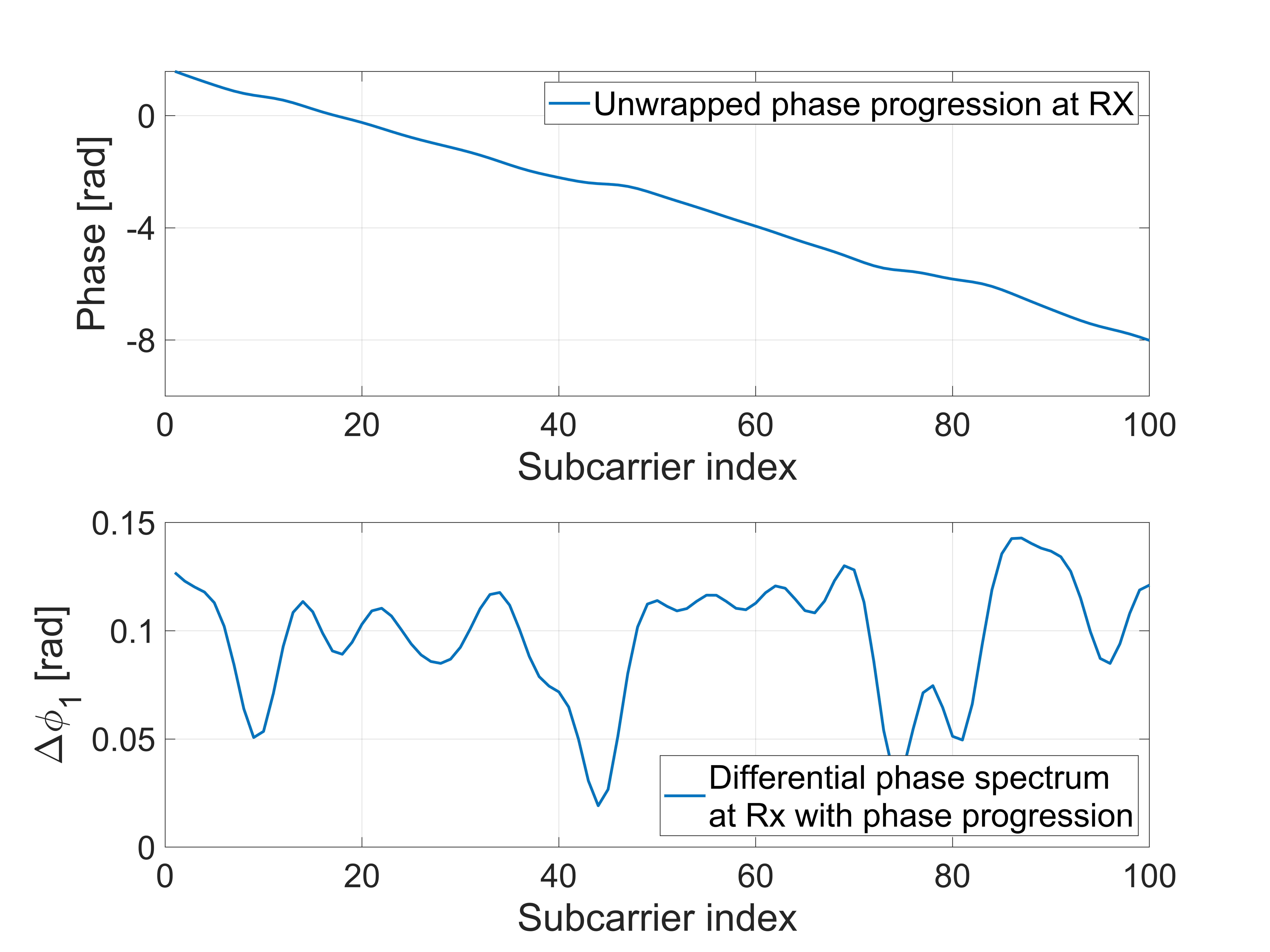}
    \vspace{-20pt}
    \caption{Differential phase spectrum at the receiver, with phase progression included.}
    \label{fig:oc_nz_pp_spectrum}
\end{figure}

}

\subsection{Simulation Scenario for 3GPP Urban Microcell (UMi)}
\label{subsec:sim-scenario}
We designed an example simulation scenario using the \textsc{QuaDriGa} framework to evaluate the CPP positioning method in a 3GPP-compliant UMi environment. The setup adheres to~\cite[Section~7.2]{3GPP_TR38_901} with the following key parameters:
\begin{itemize}
    \item Carrier frequency: 3.8\,GHz (mid-band 5G NR)
    \item Bandwidth: 100\,MHz
    \item Subcarrier spacing: 30\,kHz
    \item Antenna: Omnidirectional at TRPs and UE
\end{itemize}

\noindent Three TRPs are deployed asymmetrically with inter-site distances (ISDs) of 50--70\,m, compliant with the non-ultra-dense UMi range (50--200\,m ISD) \cite[Table~7.2-1]{3GPP_TR38_901}. The TRP locations are as follows:
\begin{itemize}
    \item TRP-1: (100, 100, 10)\,m (street-level deployment)
    \item TRP-2: (150, 90, 10)\,m 
    \item TRP-3: (140, 150, 10)\,m
\end{itemize}

\noindent \textcolor{black}{The UE is located at coordinates (120, 110, 1.5)~m and is assigned specific mobility attributes to account for low-speed urban dynamics. While the UE position remains fixed for the duration of the measurements, a velocity attribute of 3~km/h and a random waypoint motion profile are applied to the simulation parameters to account for doppler shifts.} SRS sequences are generated for 3276 subcarriers using a comb-6 configuration with 8 symbols per frame, while the channel impulse response incorporates UMi-specific delay spreads as defined in \cite[Table~7.5-6]{3GPP_TR38_901}.

The geometry-based channel model incorporates frequency-dependent path loss exponents derived from  \cite[Table~7.4.1-1]{3GPP_TR38_901}. 
The LOS probability follows the piecewise function defined in \cite[Table~7.4.2-1]{3GPP_TR38_901}.
Urban-specific effects such as clutter loss (10\,dB for street canyons) and outdoor diffraction are included to model signal attenuation in dense environments. 
The parameters align fully with \textsc{QuaDriGa}'s 3GPP UMi channel model implementation as it natively supports:
\begin{itemize}
    \item The LOS probability formula and path loss exponents from 3GPP TR~38.901~\cite{3GPP_TR38_901},
    \item Clutter loss and street canyon effects via its \texttt{clutter\_density} parameter,
    \item Delay/angular spread distributions for UMi NLOS conditions.
\end{itemize}
The asymmetric TRP layout and UE mobility model are configurable through \textsc{QuaDriGa}’s \texttt{qd\_layout} and \texttt{qd\_track} classes, ensuring reproducible results for UMi studies.
\begin{figure}[!h]
    \centering
    \includegraphics[width=0.5\textwidth]{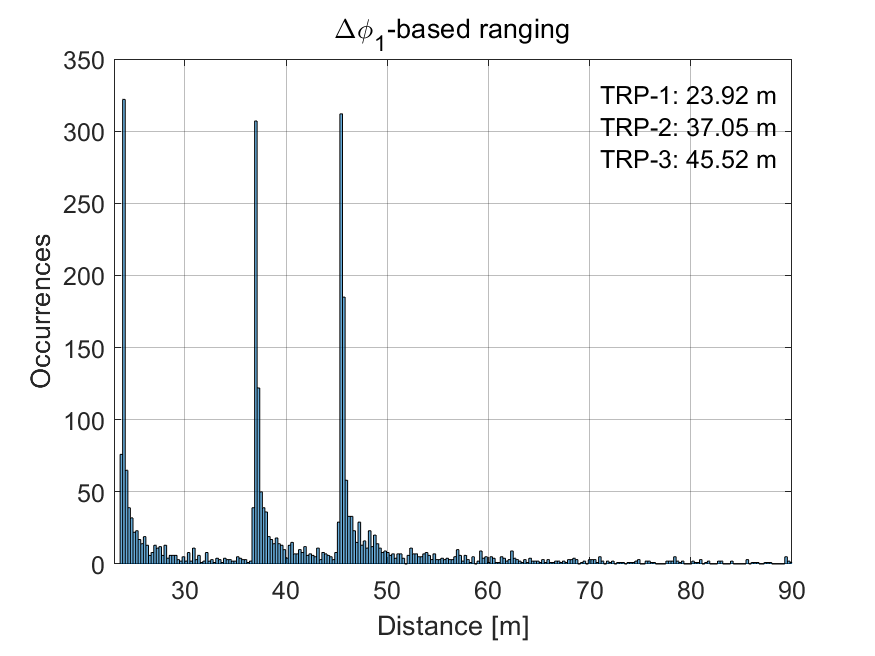}
    \caption{The Euclidean distances of the receiver to each of the transmitters.}
   \label{fig:distance_hist}
\end{figure}
The experimental evaluation, conducted over $10^3$ iterations, reveals critical insights into ranging precision through combined distance analysis (Fig.~\ref{fig:distance_hist}). The three-dimensional (3D) distance estimation to each TRP demonstrates three distinct histogram peaks at $\sim$23.9\,m, $\sim$37.1\,m, and $\sim$45.5\,m ($\pm$10\,cm resolution), corresponding precisely to simulated UE-TRP geometries. However, high-accuracy CP measurements ($\leq\,10$~cm precision) are achieved in only $\sim30\%$ of cases -- LOS conditions with high signal quality, while NLOS conditions exhibit characteristic heavy-tailed error distributions. 

Attaining centimeter-level positioning accuracy fundamentally requires simultaneous high-precision ranging to $\geq$3 TRPs under LOS propagation. This geometric constraint introduces critical performance limitations, as evidenced by the 90th percentile error reduction from 68.9\,m to 23.4\,cm in horizontal (2D) scenarios through NLOS exclusion (Table~\ref{table:pos_errors}). The positioning framework implements multi-ranging over symbols within a single transmission frame using best-measurement selection from participating TRPs, with trilateration errors constrained to $<$100\,m---a practical limit derived from the 70\,m transmitter grid spacing in our UMi deployment scenario.

\begin{table}[t]
\centering
\vspace{-8pt}
\caption{Positioning error for LOS and mixed LOS/NLOS conditions in UMi scenarios}
\label{table:pos_errors}
\begin{tabular}{l|cc|cc}
\hline
 & \multicolumn{2}{c|}{2D} & \multicolumn{2}{c}{3D} \\
\cline{2-5}
Percentile & LOS & LOS/NLOS & LOS & LOS/NLOS \\
\hline
90\% & 23.4 cm & 68.9 m & 67.5 cm & 75.3 m \\
80\% & 18.3 cm & 48.3 m & 54.4 cm & 59.1 m \\
70\% & 15.2 cm & 33.0 m & 47.1 cm & 46.5 m \\
\hline
\end{tabular}
\end{table}

The results demonstrate that even isolated NLOS errors disproportionately degrade multilateration accuracy, underscoring the critical need for robust NLOS identification and exclusion (detailed in Section~\ref{sec:coverage} and Fig.~\ref{fig:cpp_cdf}). Achieving reliable cm-level positioning in mixed environments requires two key enablers: 1) ML-assisted NLOS mitigation, and 2) geometric diversity that ensures concurrent LOS visibility to $\geq$3 spatially distributed TRPs.

\section{LOS and NLOS link classification}
\label{sec:LOS_NLOS}
\textcolor{black}{The objective of this paper is to deploy link classification in an ML-assisted positioning framework that adheres to the relevant 3GPP technical specifications and reporting requirements.} To reach this achievement, the proposed T-CNN is trained on data from the simulation setup in Sec.~\ref{subsec:sim-scenario}. The model takes the SRS waveform measured at the receiver, which reflects CIR. After FFT, a batch-normalized \textit{sequence input} layer of 3276 real power values\footnote{It corresponds to 273 PRBs, each consisting of 12 subcarriers. At 30\,kHz spacing, this yields approximately the 100\,MHz bandwidth}, is used.
 The \textit{Softmax output} layer of two neurons generates the probabilistic scores for soft decision. Since the true channel state is not directly available in \textsc{QuaDriGa}, we use the known Tx and Rx positions for labeling. True propagation delay from geometric distance is 
\begin{equation}
    \tau_{\mathrm{true}} = \frac{\| \mathbf{P}_{\mathrm{tx}} - \mathbf{P}_{\mathrm{rx}} \|}{c}\;,
    \label{tr_delay}
\end{equation}
where $c$ is the speed of light. In contrast, the estimated delay can be obtained from the PDP of a received SRS. Specifically, it corresponds to the delay associated with the strongest multipath component; thus, the absolute difference between the estimated and true delays is  
\begin{equation}
\begin{aligned}
    \tau_{\mathrm{est}} &=& \underset{\tau}{\arg\max} \,\, \text{PDP}(\tau);
\\
    \tau_{\mathrm{diff}} &=& | \tau_{\mathrm{est}} - \tau_{\mathrm{true}} |.
\end{aligned}
\label{delay_diff}
\end{equation}
The absolute delay difference is compared to a threshold of $10\,$ns—equivalent to one symbol duration given system bandwidth—to classify the SRS as LOS or NLOS. If $\tau_{\mathrm{diff}}$ exceeds the threshold, the instance is NLOS; otherwise, LOS. 

\textcolor{black}{A contribution of our proposed T-CNN is the use of \textit{stride} operation for resolution reduction, whereas many related papers rely on max-pooling as a linear summation of extracted features. Since it is a flexible sliding window mechanism for downsampling, feature map reduction is embedded in the learned convolutional operation, preserving informative multipath structure as a built-in regularizer, while reducing computation. T-CNN has the minimal architecture given our objective, as presented in Fig.~\ref{fig:tcnn_arc}, where the convolutional block (left-hand side) contains a computationally affordable convolutional layers to discover the temporal patterns caused by multipaths, and the shallow depth enables low-latency LOS/NLOS detection suitable for real-time applications. Hyperparameters were tuned for stable convergence and robust generalization by training with Adam (
$\eta_0 = 10^{-3}$) under a piecewise learning-rate schedule (drop period = 7
epochs, drop factor = 0.9), and L2 regularization ($10^{-4}$). Model selection and early stopping were driven by validation monitoring once per epoch, returning the best-validation checkpoint while optimizing precision as the objective metric. The maximum trainable parameters on the convolutional and classifier head block are 16416 and 2112, compared to 120\;K in SEL-CNN \cite{zhu}}. 
\begin{figure}[!ht]
\centering
\resizebox{1\columnwidth}{!}{%
\begin{tikzpicture}[
  node distance=5mm and 10mm,
  block/.style={draw, rounded corners, align=center, minimum height=7mm, minimum width=40mm, font=\footnotesize},
  blockw/.style={draw, rounded corners, align=center, minimum height=7mm, minimum width=44mm, font=\footnotesize},
  small/.style={draw, rounded corners, align=center, minimum height=7mm, minimum width=30mm, font=\footnotesize},
  arrow/.style={-Latex, thick}
]
\node[block] (in) {Freq.-domain input (CFR) $\mathbb{R}^{3276\times 1}$};
\node[blockw, below=6mm of in] (c1) {Conv1D (same)\\$k{=}16,\; s{=}2,\; 64$ ch $\Rightarrow \mathbb{R}^{1638\times 64}$};
\node[small, below=of c1] (bn1) {BatchNorm (64)};
\node[small, below=of bn1] (r1) {ReLU};
\node[block, below=of r1] (c2) {Conv1D (same)\\$k{=}8,\; s{=}2,\; 32$ ch $\Rightarrow \mathbb{R}^{819\times 32}$};
\node[small, below=of c2] (bn2) {BatchNorm (32)};
\node[small, below=of bn2] (r2) {ReLU};
\draw[arrow] (in) --(c1);
\draw[arrow] (c1) -- (bn1);
\draw[arrow] (bn1) -- (r1);
\draw[arrow] (r1) -- (c2);
\draw[arrow] (c2) -- (bn2);
\draw[arrow] (bn2) -- (r2);
\node[font=\footnotesize, align=left, right=2mm of c1] {Local spectral\\patterns};
\node[font=\footnotesize, align=left, right=4mm of c2] {multipath\\-induced\\selectivity};
\node[block, right=36mm of in] (gap) {Global Avg Pooling\\$\mathbb{R}^{1\times 32}$};
\node[block, below=of gap] (fc1) {FC 64};
\node[small, below=of fc1] (r3) {ReLU};
\node[block, below=of r3] (fc2) {FC 32};
\node[small, below=of fc2] (r4) {ReLU};
\node[block, below=of r4] (fc3) {FC 2};
\node[small, below=of fc3] (sm) {Softmax\\LOS / NLOS};
\draw[arrow] (gap) -- (fc1);
\draw[arrow] (fc1) -- (r3);
\draw[arrow] (r3) -- (fc2);
\draw[arrow] (fc2) -- (r4);
\draw[arrow] (r4) -- (fc3);
\draw[arrow] (fc3) -- (sm);
\draw[arrow] (r2.east) -- ++(30mm,0) |- (gap.west);
\node[draw, dashed, rounded corners, inner sep=3mm,
      fit=(c1)(r2),
      label={[font=\footnotesize, xshift=-6mm, yshift=11.7mm]above right:Feature extractor}] (leftbox) {};
\node[draw, dashed, rounded corners, inner sep=3mm,
      fit=(gap)(sm),
      label={[font=\footnotesize, yshift=15mm]above left:Classifier head}] (rightbox) {};
\end{tikzpicture}%
}
\caption{T-CNN for LOS/NLOS classification (\(L=3276\) frequency bins).}
\label{fig:tcnn_arc}
\end{figure}

\begin{figure}[!ht]
    \centering
    \includegraphics[width=1\columnwidth]{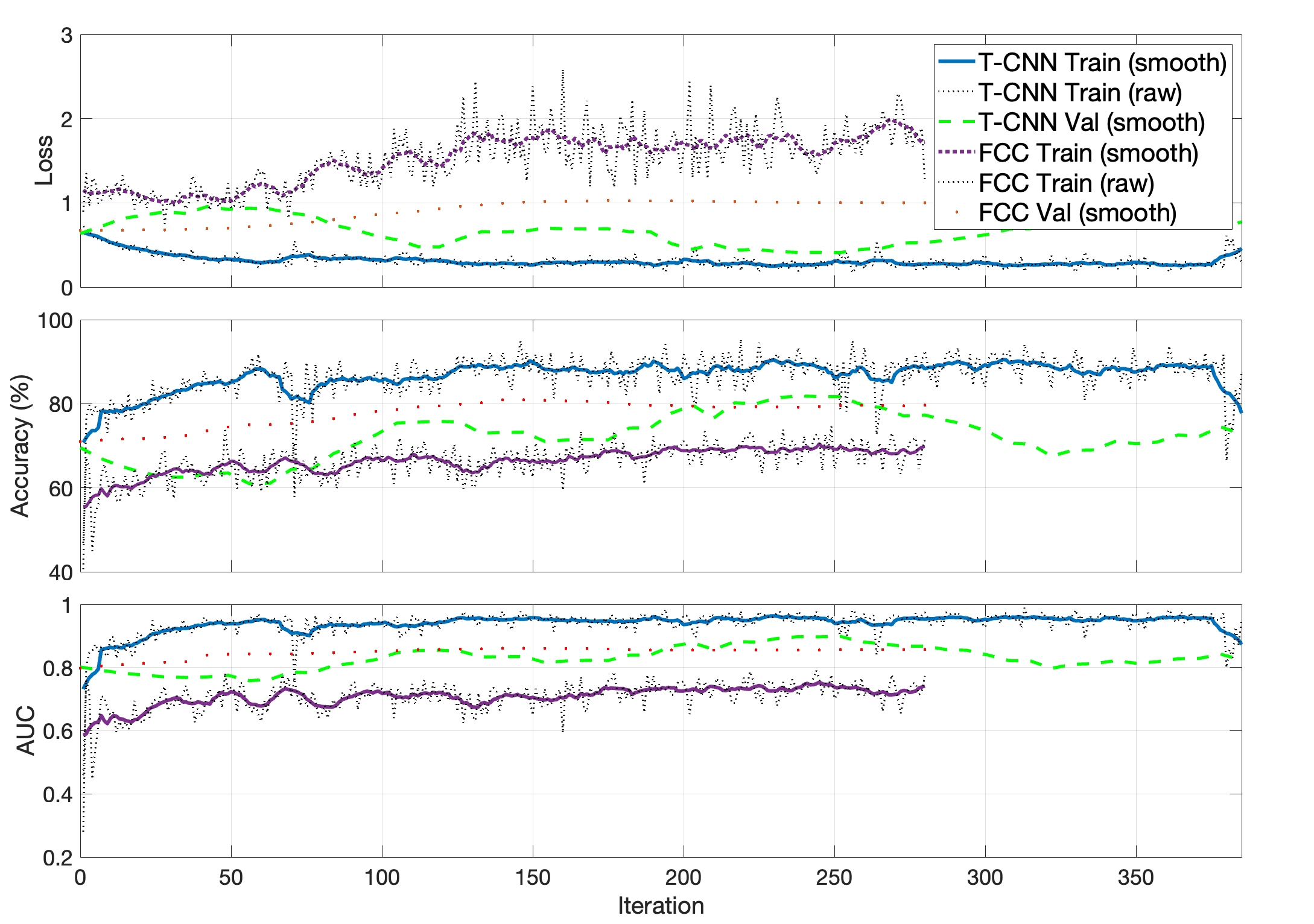}
        \vspace{-20pt}
    \caption{T-CNN vs FCC training performance, using a 10-step sliding window and an early stopping patience of 20 epochs without improvement.}
   \label{fig:dl_performce}
\end{figure}

\begin{figure}[!ht]
    \centering
    \includegraphics[width=1\columnwidth]{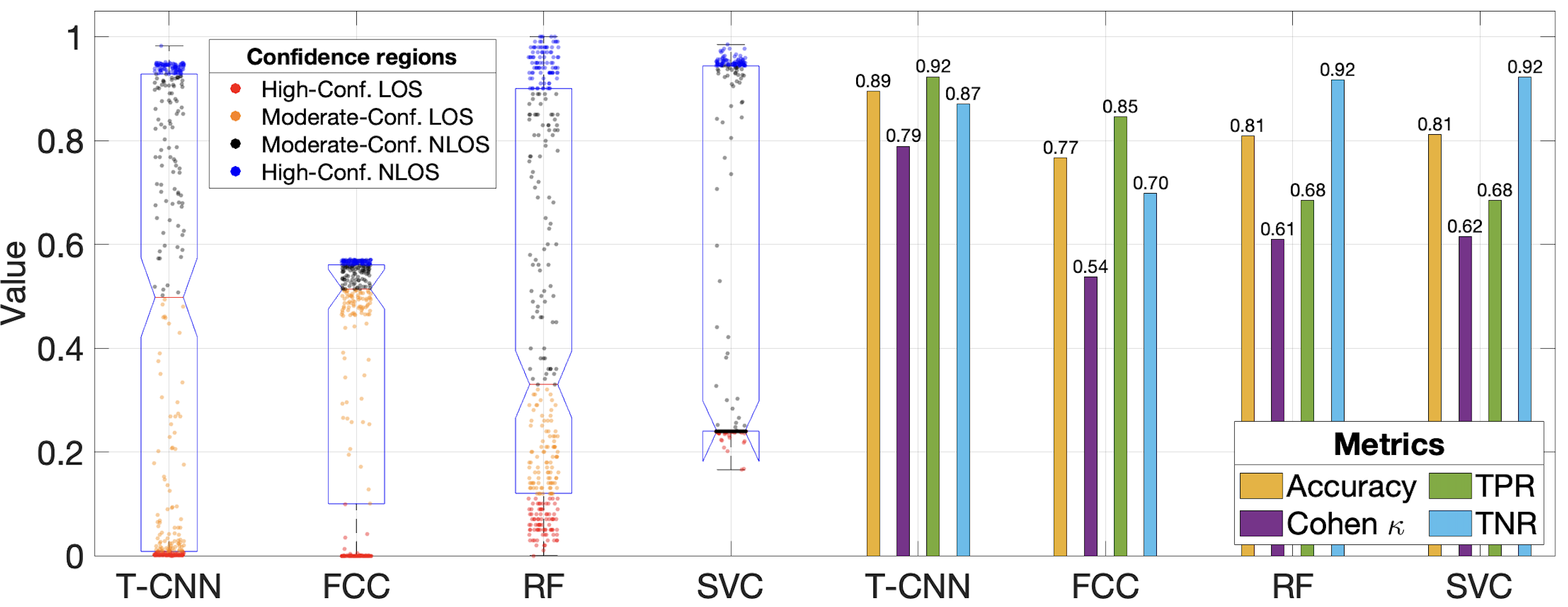}
    \vspace{-20pt}
    \caption{T-CNN prediction performance compared to other models.}
    \label{fig:pred_performance}
\end{figure}

\textcolor{black}{Training performance is illustrated in Fig.~\ref{fig:dl_performce}, and compared with a baseline Fully Connected Classifier (FCC) (right-hand block in Fig.~\ref{fig:tcnn_arc}, equivalent to a multilayer perceptron (MLP) of three hidden layers), where no feature extraction is applied. As seen, the average Area Under Curve (AUC) is 94\%, signifying that T-CNN has an excellent class separability performance on a broad range of threshold values. T-CNN accuracy is 89\% on average, compared with 77\% for FCC. Additionally, a Support Vector classifier (SVC) with an \textit{Radial Basis Kernel Functions (RBF)} and a Random Forest (RF) classifier of 100 trees with \textit{bagging} method is also trained as a basis for comparison. Their prediction performance is presented in Fig.~\ref{fig:pred_performance}. As seen, the accuracy of all classifiers (except FCC) is above 80\% while SVC and RF have a bias toward the LOS true prediction and FCC has a bias toward NLOS true prediction. Since classification accuracy alone is misleading, we additionally report Cohen’s $\kappa$ score, which quantifies agreement beyond chance as 
\begin{equation}
    \kappa = \frac{P_o - P_e}{1 - P_e}, 
\end{equation}
where $P_o$ is the observed accuracy (empirical agreement) and $P_e$ is the expected agreement under chance given the marginal class frequencies. $P_e$ is close to 0.5 for all classifiers, indicating that the label proportions are balanced. Additionally, the most substantial $\kappa$ score (see \cite{kabir2021csi} for qualitative interpretation of $\kappa$) belongs to T-CNN, i.e., the T-CNN decisions are strongly consistent with the ground truth after accounting for the agreement expected from the class marginals.} 

\begin{figure}[h]
    \centering
    \includegraphics[width=0.5\textwidth]{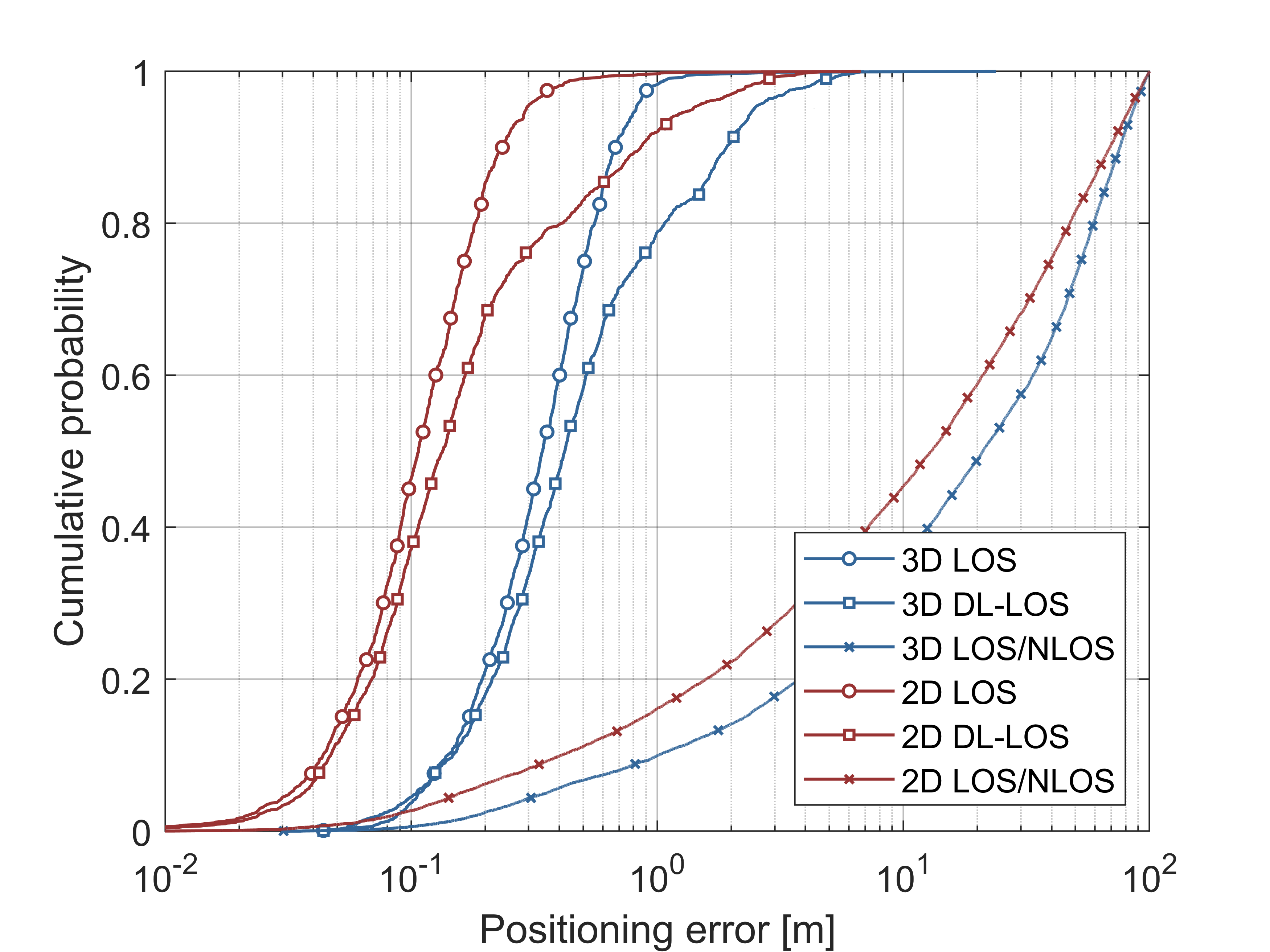}
    \vspace{-20pt}
    \caption{CDF of 2D/3D positioning errors under LOS-only, mixed LOS/NLOS conditions, and T-CNN assisted LOS identification in UMi scenarios.}
   \label{fig:cpp_cdf}
\end{figure}

\textcolor{black}{Fig.~\ref{fig:pred_performance} also summarizes the confidence regions for each classifier. Given the softmax outputs, we compute a confidence measure $\max(p_{\mathrm{LOS}},p_{\mathrm{NLOS}})$, where $p(.)$ is the label probabilities. Consequently, a soft confidence-aware prediction given each classifier is presented in addition to hard label prediction. To segment the regions, a boxplot and Inter-Quantile-Range (IQR) analysis is applied and visualized. IQR range of 25\% to 50\% are selected for moderate confidence range; a prediction has high confidence outside this range. As expected, the dense regions are close to zero and one values and the median is 0.5 for T-CNN. However, setting an ordinary threshold of 0.5 for hard decision, yields a moderately confident labeling for SVC and RF. Additionally, the FCC prediction is not confident since it does not cover all the probabilistic values, i.e., the highest confident value for NLOS prediction is below 0.6.}     
To compare the impact of our NLOS detection-and-discard approach by using T-CNN, we evaluate the downstream positioning accuracy with and without NLOS links, in Fig.~\ref{fig:cpp_cdf}, where the cumulative distribution function (CDF) of 2D and 3D positioning errors across are presented in three scenarios: (i) only LOS links, (ii) a mix of LOS and NLOS links, and (iii) a mixed scenario where the T-CNN model is used to identify and exclude NLOS links through a hard decision. The results show that using T-CNN NLOS filtering significantly reduces positioning error, bringing the performance close to the pure LOS case.
Simulation results show that the our approach improves the quality of CP measurements and enhances positioning accuracy in complex environments.

\section{Sensor Fusion with Integrated Carrier Phase Positioning}
\label{sec:sensor_fusion}

To ensure robust and high-precision localization under dynamic motion, variable visibility, and LOS/NLOS channel transitions, we integrate information from multiple complementary sensors. Specifically, we integrate inertial measurement data and stereo visual odometry with a CPP-based positioning technique. The fusion framework is implemented using an ES-EKF, which provides a principled probabilistic framework for integrating sensor data with different noise characteristics and availability, including visual odometry, inertial measurements, and CPP updates that may be intermittent due to varying LOS/NLOS conditions.

\subsection{Visual Odometry (VO)}
VO is a method that uses sequential image data from a camera sensor to compute and estimate motion. The motion of the agent is tracked across multiple  sequential frames, essentially providing information on the agent's trajectory. The VO system used in this work is a feature-based stereo-visual odometry. Knowing the starting position, we can then track the agent's motion across the map. By the application of RANSAC (RANdom SAmple Consensus) based outlier detection system, the influence of dynamic moving objects on the computation is reduced, and the reliability of the visual odometry estimates is increased. 
This is because the estimate of the motion is highly susceptible to the environment. 
This susceptibility arises from the fact that dynamic elements in the environment can corrupt the motion estimation by introducing extraneous motion components that are incorrectly attributed to the agent.

\subsection{Inertial Odometry}

Inertial sensors can measure the acceleration and angular velocity of an agent. However, estimating motion solely based on these measurements is highly error-prone due to sensor bias, noise, and integration drift over time. Although displacement can be derived from velocity and time using the basic laws of motion (i.e., kinematic equations), such calculations are highly sensitive to sensor errors, making them unreliable when relying on inertial sensors alone. Despite these limitations, inertial data remains valuable in sensor fusion systems. Over short time intervals, motion estimates derived from inertial sensors are sufficiently accurate to provide useful complementary information that enhances the overall robustness and precision of fusion-based motion estimation.

\subsection{Carrier Phase Positioning Integration}
\textcolor{black}{The CPP module provides absolute position estimates obtained via trilateration from carrier-phase-based range measurements to at least three LOS base stations. The CPP-based trilateration is performed independently, and the resulting 3D position estimate is injected into the ES-EKF as a direct measurement update.
The CPP update is triggered only when at least three base stations satisfy LOS conditions, as determined by the ML-based classifier described in Section~\ref{sec:LOS_NLOS}. In the absence of sufficient LOS connectivity, the filter relies solely on IMU propagation and visual odometry updates.
The covariance matrix is set based on empirical error statistics observed in Section~\ref{subsec:sim-scenario} (and can be tuned per deployment), reflecting centimeter-level variance under LOS conditions. This adaptive measurement modeling ensures appropriate weighting between CPP, visual odometry, and inertial propagation within the Kalman update step.}

\subsection{Error State Extended Kalman Filter}
The proposed ES-EKF filter estimates the agent's state, consisting of the 3D position $\mathbf{p} \in \mathbb{R}^3$, velocity $\mathbf{v} \in \mathbb{R}^3$, and orientation $\mathbf{q} \in \mathbb{R}^4$. The state vector at time step $k$ is defined as $\mathbf{x}_k = [\mathbf{p}_k^\mathrm{T},  \mathbf{v}_k^\mathrm{T}, \mathbf{q}_k^\mathrm{T}]^\mathrm{T} \in \mathbb{R}^{10}$. Using the nonlinear motion model, we propagate/predict the nominal state as:
\begin{equation}
\begin{aligned}
\mathbf{p}_k = \mathbf{p}_{k-1} + \Delta t \cdot \mathbf{v}_{k-1}
      + \frac{\Delta t^2}{2} \cdot \left( \mathbf{R}(\mathbf{q}_{k-1}) \cdot \mathbf{f}_{k-1} + \mathbf{g} \right)
\end{aligned}
\end{equation}
\vspace{-10pt}
\begin{equation}
\begin{aligned}
\mathbf{v}_k = \mathbf{v}_{k-1} + \Delta t \cdot \left( \mathbf{R}(\mathbf{q}_{k-1}) \cdot \mathbf{f}_{k-1} + \mathbf{g} \right)
\end{aligned}
\end{equation}
\begin{equation}
\mathbf{q}_k = \mathbf{q}_{k-1} \otimes \mathbf{q}(\boldsymbol{\omega}_{k-1} \, \Delta t)
\label{eqn:quat_motion_model}
\end{equation}
where $\mathbf{p}_{k-1}$, $\mathbf{v}_{k-1}$, and $\mathbf{q}_{k-1}$ represent the previous position, velocity, and orientation states, respectively. $\Delta t$ represents the time between state estimates, and $\mathbf{g} \in \mathbb{R}^3$ is the gravity vector. The symbol $\otimes$ in  ~\eqref{eqn:quat_motion_model} denotes the quaternion multiplication and $\mathbf{q}(\boldsymbol{\omega}_{k-1} \, \Delta t)$ converts the Euler angles $\boldsymbol{\omega}$ to quaternion representation. $\mathbf{f}_k \in \mathbb{R}^3$ and $\mathbf{w}_k \in \mathbb{R}^3$ are the linear acceleration and angular velocity measured by the IMU. $\mathbf{q}_{k-1}$ is the orientation quaternion at time $k-1$, and $\mathbf{R}(\mathbf{q}_{k-1})$ is the corresponding rotation matrix that transforms vectors from the body (IMU) frame to the world (inertial) frame.

To robustly fuse sensor data with different update rates and to mitigate drift and noise—especially from inertial measurements—we use the \textit{error-state} formulation of the EKF. Unlike the classical EKF, which directly estimates the full state, the ES-EKF maintains a nominal state and estimates a small additive error state. This approach improves numerical stability for orientation (notably for quaternions) and simplifies linearization by applying it only to the error dynamics rather than the full nonlinear model. The nominal state is propagated using the nonlinear model above, while the error state is handled using a linearized system with Gaussian noise. The following section details this process.
\subsubsection{Error State Model}
The error state represents small deviations from the nominal state:
\begin{equation}
\mathbf{\delta x_k} \;=\;
\begin{bmatrix}
\delta \mathbf{p}_k\\[6pt]
\delta \mathbf{v}_k\\[6pt]
\delta \boldsymbol{\theta}_k
\end{bmatrix}
\;\in\;\mathbb{R}^9
\end{equation}
The error dynamics evolve as a linear time-varying system
\begin{equation}
\mathbf{\delta x_k} \;=\; \mathbf{F}_{k-1}.\,\delta \mathbf{x}_{k-1} \;+\; \mathbf{L}_{k-1}. \,\mathbf{n}_{k-1}
\end{equation}
where $\mathbf{F}_{k-1} \in \mathbb{R}^{9 \times 9}$ is the state transition matrix of the error-state dynamics with respect to the error state itself, and  $\mathbf{L}_{k-1} \in \mathbb{R}^{9 \times 6}$ is the noise mapping matrix that maps the IMU process noise into the error-state space, both evaluated at time step $k-1$. 
\begin{equation}
\label{eqn:fk_1}
\mathbf{F}_{k-1} \;=\;
\begin{bmatrix}
\mathbf{I}_3 & \Delta t\,\mathbf{I}_3 & \mathbf{0}_{3\times3} \\[6pt]
\mathbf{0}_{3\times3} & \mathbf{I}_3 & - \mathbf{R}(\mathbf{q}_{k-1}) \cdot \mathbf{f}_{k-1} \times \Delta t \\[6pt]
\mathbf{0}_{3\times3} & \mathbf{0}_{3\times3} & \mathbf{I}_3,
\end{bmatrix},
\end{equation}

\begin{equation}
\label{eqn:lk_1}
\quad
\mathbf{L}_{k-1} \;=\;
\begin{bmatrix}
\mathbf{0}_{3\times3} & \mathbf{0}_{3\times3}\\[6pt]
\mathbf{I}_3 & \mathbf{0}_{3\times3}\\
\mathbf{0}_{3\times3} & \mathbf{I}_3
\end{bmatrix},
\end{equation}
and $\mathbf{{I}_3}$ represents a $3\times3$ identity matrix.\\

\subsubsection{Prediction Step Noise}

The prediction step is affected by noise $\mathbf{n_{k}} \;\sim\; \mathcal{N}(\mathbf{0},\,\mathbf{Q}_k)$ where

\begin{equation}
\mathbf{Q_k} \;=\; \Delta t^2
\begin{bmatrix}
\boldsymbol{\sigma}_{\mathrm{acc}}^2 & \mathbf{0}_{3 \times 3}\\[6pt]
\mathbf{0}_{3 \times 3} & \boldsymbol{\sigma}_{\mathrm{gyr}}^2
\end{bmatrix}
\end{equation}
where $ {\sigma}_{\mathrm{acc}}^2 $  and $ {\sigma}_{\mathrm{gyr}}^2 $ are the variances of the accelometer and the gyroscope.

\subsubsection{Observation Noise}

The observation noise is modelled as $\mathbf{o}_k \sim \mathcal{N}(\mathbf{0},\,\mathbf{R}_k)$
where $\mathbf{R}_k \in \mathbb{R}^{3 \times 3}$ is the measurement noise covariance matrix, representing the uncertainty in either the visual odometry or the CPP-based position measurement, depending on which observation is used at time step $k$.
$\mathbf{R}_k$ is used in the measurement update step of the Kalman filter. Specifically, it appears in the computation of the Kalman gain, where it quantifies the expected uncertainty in the incoming position measurement.

\begin{figure*}[th!]
    \centering
    \includegraphics[width=0.8\textwidth]{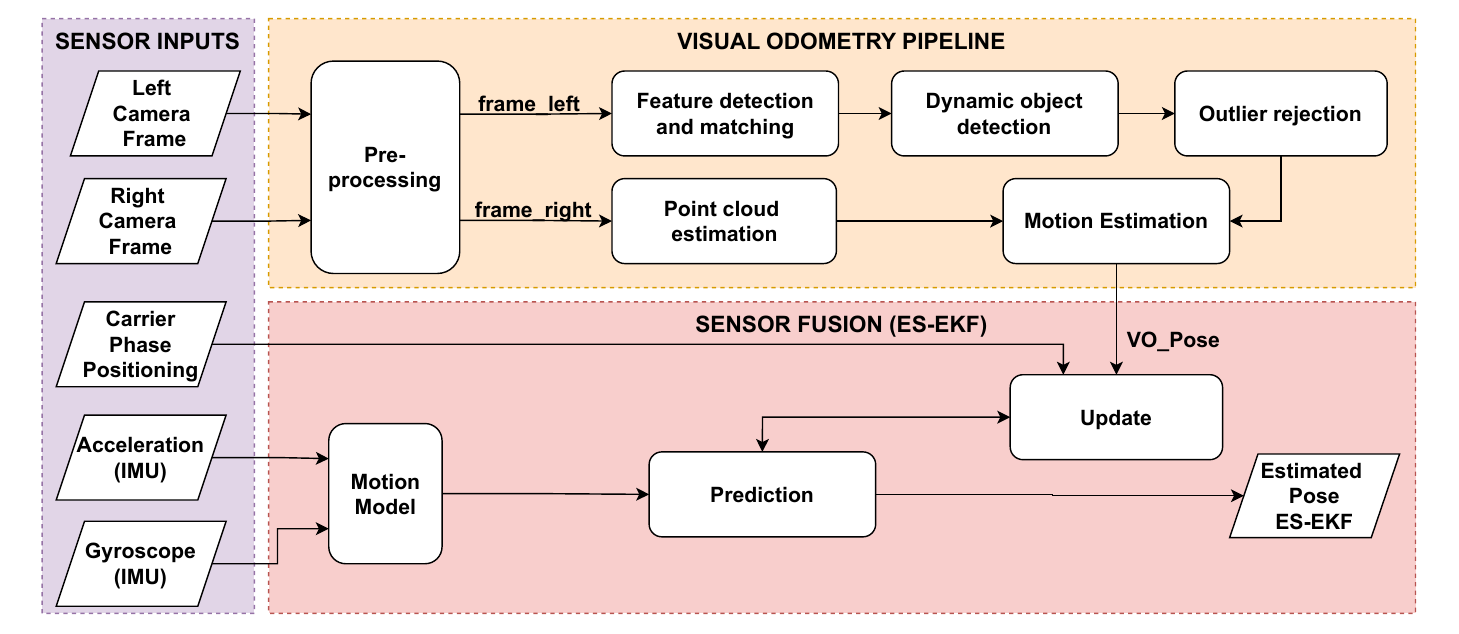}  
    \caption{ES-EKF--based sensor fusion framework integrating 5G carrier-phase positioning, IMU measurements, and stereo visual odometry for robust pose estimation.}
   \label{fig:esekf_vo_diagram}
\end{figure*}

Based on the above modeling, we formulate the fusion algorithm shown in Algorithm~\ref{alg:es_ekf_update}, which integrates information from inertial, visual, and CPP-based positioning. Note that variables with a check mark (ˇ) denote predicted values from the motion model (prior estimates), while variables with a hat (ˆ) represent corrected values after incorporating measurements (posterior estimates). The overall system architecture is shown in Fig.~\ref{fig:esekf_vo_diagram}.

\begin{algorithm}[htbp]
  \caption{ES-EKF based Fusion of VO, IMU and CPP Measurements}
  \label{alg:es_ekf_update}
  \begin{algorithmic}[1]
    \REQUIRE Previous state $(\check{\mathbf{p}}_{k-1},\,\check{\mathbf{v}}_{k-1},\,\check{\mathbf{q}}_{k-1})$, 
             covariance $\mathbf{P}_{k-1}$, IMU acceleration $\mathbf{f}_{k-1}$, angular velocity $\boldsymbol{\omega}_{k-1}$,  
             process noise $\mathbf{Q}_{k-1}$, measurement noise covariance $\mathbf{R_k}$,  
             measurement matrix $\mathbf{H}_k$,  
             measurement $\mathbf{y}_k = \mathbf{H}_k x_k + \mathbf{o}_k$ (if available, here $\mathbf{y}_k$ is from either visual odometry or CPP)
    \ENSURE Updated state $(\hat{\mathbf{p}}_k,\,\hat{\mathbf{v}}_k,\,\hat{\mathbf{q}}_k)$ and covariance $\mathbf{P}_k$

    \STATE \textbf{Predict state:}
    \STATE\quad $\check{\mathbf{p}}_k \leftarrow \check{\mathbf{p}}_{k-1} + \Delta t\,\check{\mathbf{v}}_{k-1} + \frac{1}{2}\Delta t^2\,\left(\mathbf{R}(\check{\mathbf{q}}_{k-1}) \cdot \mathbf{f}_{k-1} + \mathbf{g} \right)
$ \hfill \texttt{\footnotesize \# Eq.~(9)}
    \STATE\quad $\check{\mathbf{v}}_k \leftarrow \check{\mathbf{v}}_{k-1} + \Delta t\,(\mathbf{f}_{k-1} + \mathbf{g})$ \hfill \texttt{\footnotesize \# Eq.~(10)}
    \STATE\quad $\check{\mathbf{q}}_k \leftarrow \check{\mathbf{q}}_{k-1} \otimes \mathbf{q}(\boldsymbol{\omega}_{k-1}\Delta t)$ \hfill \texttt{\footnotesize \# Eq.~(11)}

    \STATE \textbf{Propagate covariance:}
    \STATE\quad $\check{\mathbf{P}}_k \leftarrow \mathbf{F}_{k-1}\,\mathbf{P}_{k-1}\,\mathbf{F}_{k-1}^\top \;+\; \mathbf{L}_{k-1}\,\mathbf{Q}_{k-1}\,\mathbf{L}_{k-1}^\top$

    \IF{measurement $\mathbf{y}_k$ is available}
      \STATE \textbf{Compute Kalman gain:}
      \STATE\quad $\mathbf{K}_k \leftarrow \check{\mathbf{P}}_k\,\mathbf{H}_k^\top\bigl(\mathbf{H}_k\,\check{\mathbf{P}}_k\,\mathbf{H}_k^\top + \mathbf{R_k}\bigr)^{-1}$
      \STATE \textbf{Estimate error state:}
      \STATE\quad $\delta \mathbf{x}_k \leftarrow \mathbf{K}_k\,\bigl(\mathbf{y}_k - h(\check{\mathbf{x}}_k)\bigr)$, here $ h(\check{\mathbf{x}}_k) $ is $ \check{\mathbf{P}}_k$, assuming that the observation directly measures position.
      \STATE \textbf{Correct predicted state:}
      \STATE\quad $\hat{\mathbf{p}}_k \leftarrow \check{\mathbf{p}}_k + \delta \mathbf{p}_k$
      \STATE\quad $\hat{\mathbf{v}}_k \leftarrow \check{\mathbf{v}}_k + \delta \mathbf{v}_k$
      \STATE\quad $\hat{\mathbf{q}}_k \leftarrow \check{\mathbf{q}}_k \otimes \mathbf{q}(\delta \boldsymbol{\theta}_k)$
      \STATE \textbf{Update covariance:}
      \STATE\quad $\mathbf{P}_k \leftarrow (\mathbf{I}_{9 \times 9} - \mathbf{K}_k\,\mathbf{H}_k)\,\check{\mathbf{P}}_k$
    \ELSE
      \STATE \textbf{No available measurement from either visual odometry or CPP:}
      \STATE\quad $(\hat{\mathbf{p}}_k,\,\hat{\mathbf{v}}_k,\,\hat{\mathbf{q}}_k,\,\mathbf{P}_k) \leftarrow (\check{\mathbf{p}}_k,\,\check{\mathbf{v}}_k,\,\check{\mathbf{q}}_k,\,\check{\mathbf{P}}_k)$
    \ENDIF

    \RETURN $(\hat{\mathbf{p}}_k,\,\hat{\mathbf{v}}_k,\,\hat{\mathbf{q}}_k,\,\mathbf{P}_k)$
  \end{algorithmic}
\end{algorithm}

\subsection{Scenario of fusion experiment}

Let's consider a typical scenario where a vehicle is an agent whose motion is being estimated using an integrated CP-based positioning system, stereo visual odometry, and inertial sensors. All individual source of sensor information have their own shortcomings. For example, the integrated CPP system is susceptible to the number of BSs available and LOS conditions during the estimation of the location. The application of the fusion should rectify this issue in NLOS conditions, and the inertial sensor would augment the visual odometry in estimating the trajectory.

\begin{figure}[ht!]
    \centering
    \includegraphics[width=0.4\textwidth]{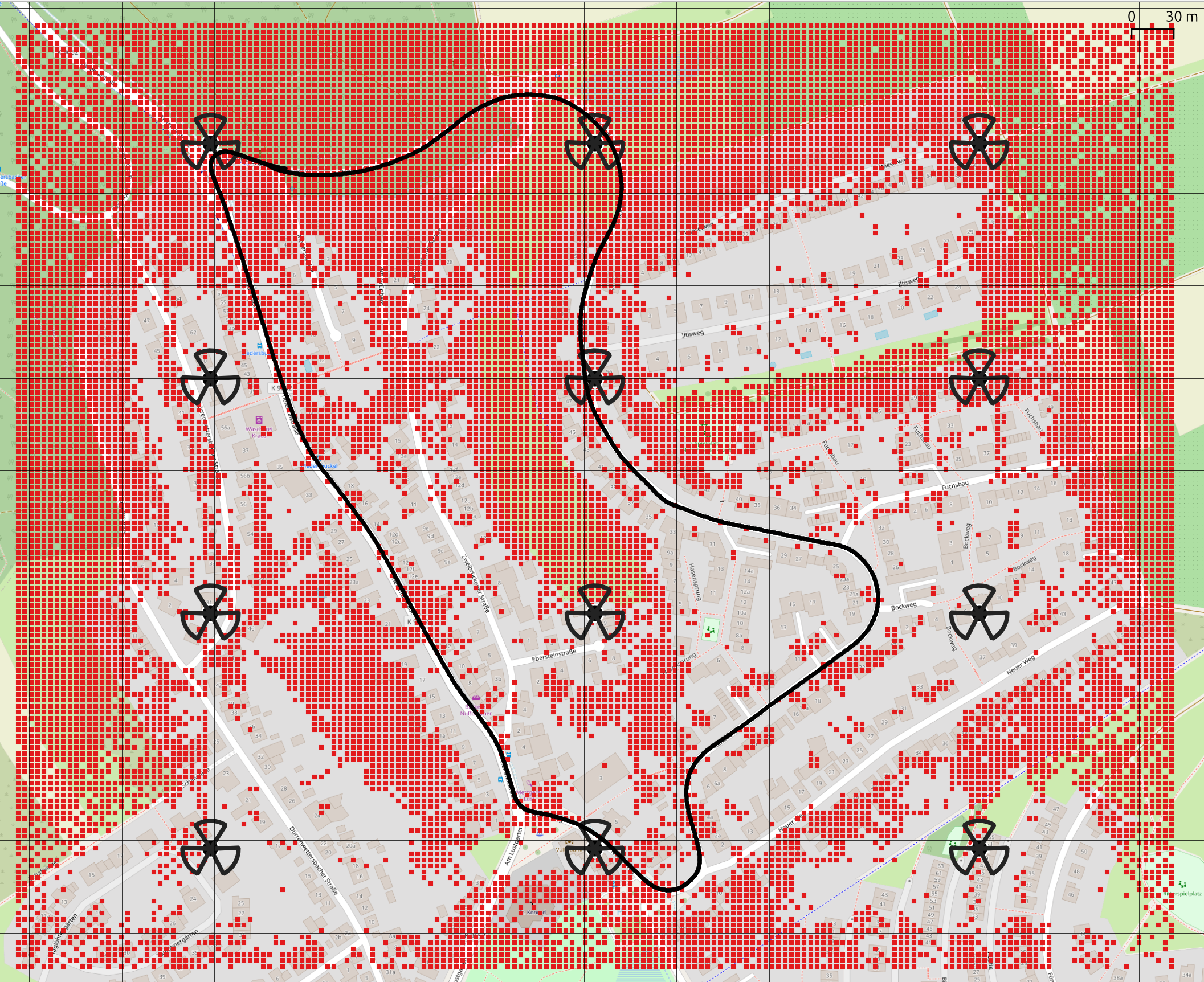}  
    \caption{Sequence 09 given by the black trajectory overlayed on the LOS/NLOS data obtained using simulation. In this figure, red squares indicate locations where at least three BSs maintain LOS conditions.}
   \label{fig:los_nlos_simulation_seq_09}
\end{figure}

For this scenario, we utilize the KITTI dataset \cite{kitti}. Specifically, we use sequence 09 of the KITTI dataset to validate our proposed fusion system. The KITTI dataset provides recorded information on camera frames, IMU, and GPS data of a vehicle moving through an urban street. The GPS and IMU data is collected using OXTS sensor, which uses RTK to provide precise ground truth information on the trajectory. This is necessary to compare our computed trajectory with the ground truth trajectory. We simulate SRS and PRS-based trilateration with the LOS/NLOS data obtained from the simulation conducted using the methods described in Sec.~\ref{sec:coverage}. Fig.~\ref{fig:los_nlos_simulation_seq_09} visualizes the trajectory of sequence 09 using the ground truth information overlayed by the simulated LOS/NLOS paths for the same area in which sequence 09 is recorded.

\begin{figure}[ht!]
    \centering
    \includegraphics[trim={0 1cm 0 2.3cm},clip, width=0.5\textwidth]{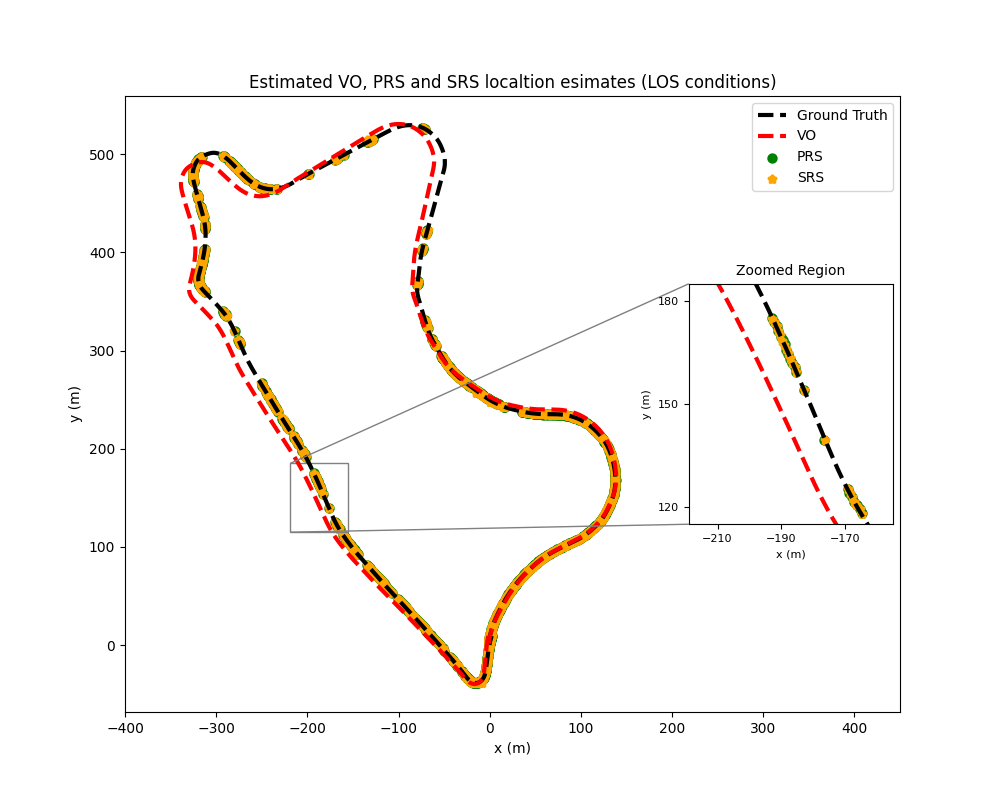}
    \vspace{-15pt}
    \caption{Visual odometry, SRS, and PRS-based trajectory estimates on sequence 09. The zoomed region shows the inconsistent trajectory estimate by SRS and PRS due to the NLOS issue.}
\label{fig:compare_vo_prs_srs_trajectories}
\vspace{-10pt}
\end{figure}

\begin{figure}[t!]
    \centering
    \includegraphics[trim={0 1cm 0 2.35cm},clip, width=0.5\textwidth]{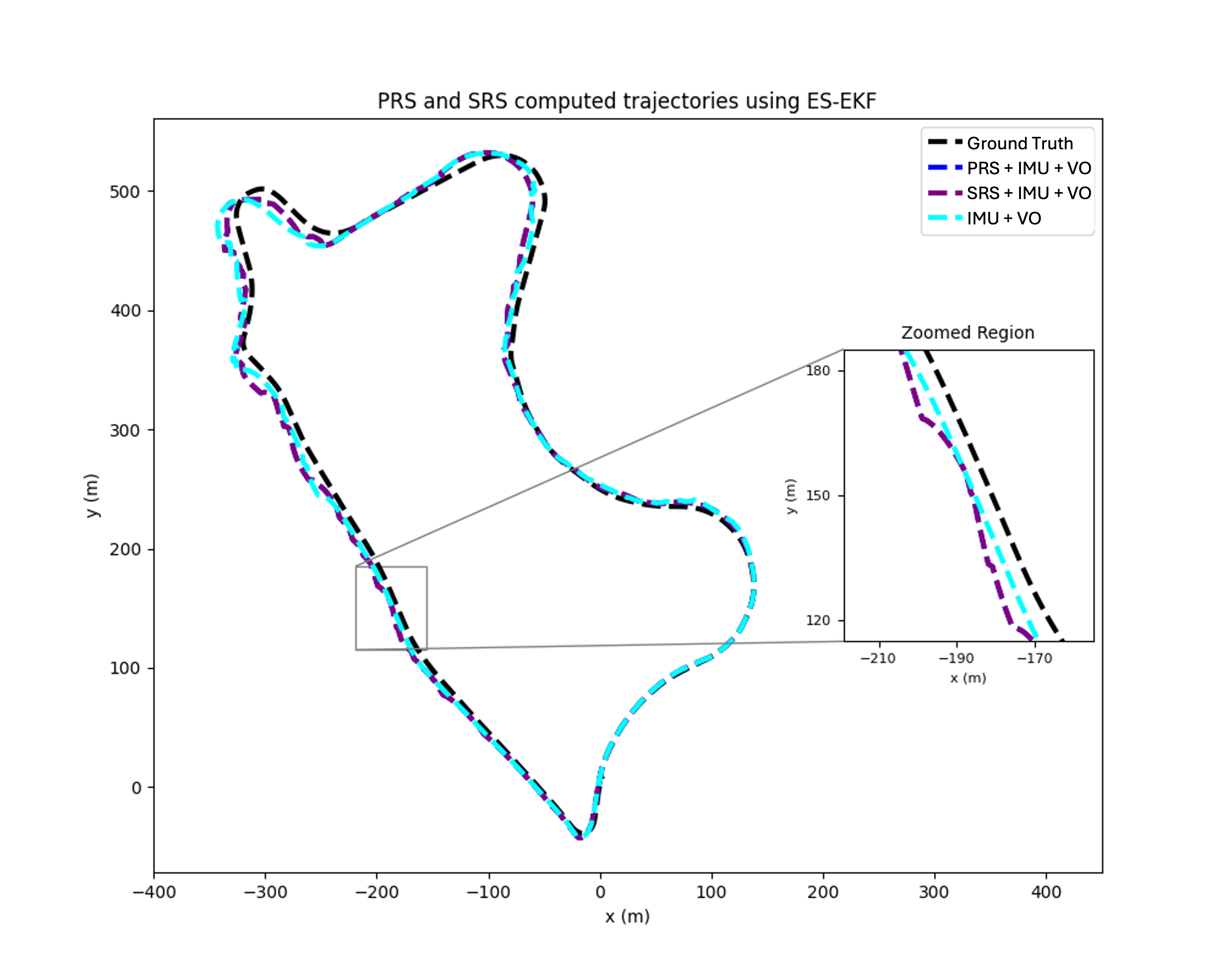}
    \vspace{-15pt}
    \caption{Estimated trajectories using ES-EKF for both SRS and PRS based trajectory estimation.}
   \label{fig:ekf_trajectories}
   \vspace{-10pt}
\end{figure}

Figure~\ref{fig:compare_vo_prs_srs_trajectories} compares the trajectory estimates from visual odometry and CPP using SRS and PRS signals. The CPP-based trajectories are shown only at locations where at least three BSs provide LOS conditions. Notably, when this condition is met, the CPP trajectory closely aligns with the ground truth, demonstrating the high positioning accuracy of the CPP method.
To obtain a complete and consistent trajectory, we employ the proposed fusion system to integrate IMU and VO data with CPP. The resulting trajectory is shown in Fig.~\ref{fig:ekf_trajectories}.

Figure~\ref{fig:ekf_trajectories} presents a limited set of trajectory points to provide a visual overview of the estimated path; however, this is not sufficient for drawing definitive conclusions about the system's performance.
To quantitatively evaluate the accuracy of the estimated trajectories, we use 
the absolute trajectory error (ATE) and relative pose error (RPE) metrics, as defined in \eqref{eqn:ate} and \eqref{eqn:rpe}, respectively. ATE measures the square root of the average of the squared Euclidean distances between the estimated positions and the ground truth positions over all $N$ timestamps or positions in the trajectory. 
\(P^{est}_{k}\) represents the estimated trajectory and \(P^{gt}_{k}\) represents the ground truth. Meanwhile, average RPE (translational) measures error in translation over a fixed interval $\Delta$. \(P_{t}\) and \(P_{t+\Delta}\) are pose 
estimates at time \(t\) and \(t+\Delta\). \(Q_{t}\) and \(Q_{t+\Delta}\) are corresponding poses in the ground truth trajectory. \((P_{t} P_{t+\Delta})_{\mathrm{trans}}\) and \((Q_{t} Q_{t+\Delta})_{\mathrm{trans}}\) are the translational components of the estimated and ground truth poses, respectively. 
\begin{equation}
\label{eqn:ate}
\mathrm{ATE} = \sqrt{
      \sum_{k=1}^{N}
        \bigl\lVert p_{k}^{est} - p_{k}^{gt}\bigr\rVert^{2}
    } \,
\end{equation}
\begin{equation}
\label{eqn:rpe}
\mathrm{RPE}_{\mathrm{trans}} \!=\! \frac{1}{\,N \!- \!\Delta\,}\!\!
   \sum_{t=1}^{\,N - \Delta\,}\!
     \bigl\lVert
       \bigl(Q_{t}^{-1}Q_{t+\Delta}\bigr)_{\mathrm{trans}}
       \;\!\!\!\!\!-\!\;
       \bigl(P_{t}^{-1}P_{t+\Delta}\bigr)_{\mathrm{trans}}
     \bigr\rVert
\end{equation}
\vspace{-8pt}
\begin{table}[htbp]
  \centering
  \vspace{-8pt}
  \caption{Absolute Trajectory Error from ES‐EKF}
  \label{tab:ate_es_ekf}
  \begin{tabular}{lcc}
    \toprule
    \textbf{Sensors Fused} & \textbf{ATE (m)} & \textbf{RPE (deg)} \\
    \midrule
    VO           & 11.56            & 0.081               \\
    IMU + VO          & 15.60            & 0.068               \\
    SRS + IMU + VO           & \textbf{4.63}             & 0.082               \\
    PRS + IMU + VO           & \textbf{4.66}             & 0.081               \\
    \bottomrule
  \end{tabular}
  \label{tabel:es-ekf-error}
  \vspace{-5pt}
\end{table}

From Table~\ref{tabel:es-ekf-error}, we can deduce that the integration of both SRS and PRS-based position estimates significantly improves the overall estimated error.
Comparing our results with those from \cite[Table 1]{dfvo-vio}, which meticulously collected data from other state-of-the-art systems, we find that our proposed solution achieves a marked improvement in ATE over all other state-of-the-art systems, by a factor of 2.

\section{Conclusions}
\label{sec:conclusion}
This paper presents a robust positioning system that leverages 5G infrastructure as a terrestrial alternative to traditional GNSS-based localization. Motivated by the challenges of operating in GNSS-denied scenarios, the proposed solution integrates Carrier Phase Positioning with a machine learning-based LOS and NLOS wireless channel classification technique and multi-sensor fusion (using IMU and visual odometry data) into a unified, resilient positioning system. 

We address key challenges in modern positioning systems, including carrier phase ambiguity resolution, measurement errors due to multipath effects, and positioning continuity during LOS outages. Our multifrequency-based phase averaging approach enables single-epoch range estimation, simplifying system architecture while maintaining centimeter-level accuracy in LOS condition. Furthermore, we propose a deep learning framework for real-time LOS/NLOS classification using SRS-derived channel impulse responses, which significantly improves positioning robustness by classifying and discarding corrupted NLOS signals. The integration of visual odometry and inertial data with cellular sounding reference signal within an Error-State Extended Kalman Filter enables seamless trajectory tracking, even under conditions where a single positioning measurement or technology becomes degraded or unavailable.

Through extensive simulations based on 3GPP-compliant channel models and real-world evaluation using the KITTI dataset, we demonstrate that our system achieves high positioning accuracy and continuity across varied scenarios. Experimental results confirm the system’s effectiveness in maintaining localization with absolute trajectory errors below 5 meters, even in environments with significant signal obstructions. Overall, this study validates the feasibility of using 5G networks as a resilient, scalable, and high-accuracy positioning platform, offering a promising path forward for mission-critical applications. \textcolor{black}{Our KITTI experiment uses a deployment-aware hybrid setup in which the 5G CPP measurement stream is simulated, since CPP-grade PRS/SRS carrier-phase measurements synchronized with camera/IMU and high-precision ground truth are currently unavailable in public datasets and are difficult to obtain in reproducible multi-gNB field trials. Absolute errors may vary in real deployments due to gNB synchronization quality, RF impairments, and PRS/SRS scheduling policies. Therefore, future work will focus on real-world deployments on a controllable multi-gNB setup, optimization of BS placement for urban and indoor scenarios, and the extension of CPP methods to support high-mobility users.}



\bibliographystyle{IEEEtran}
\bibliography{IEEEabrv, MyBib}

\begin{thebibliography}{10}
\providecommand{\url}[1]{#1}
\csname url@samestyle\endcsname
\providecommand{\newblock}{\relax}
\providecommand{\bibinfo}[2]{#2}
\providecommand{\BIBentrySTDinterwordspacing}{\spaceskip=0pt\relax}
\providecommand{\BIBentryALTinterwordstretchfactor}{4}
\providecommand{\BIBentryALTinterwordspacing}{\spaceskip=\fontdimen2\font plus
\BIBentryALTinterwordstretchfactor\fontdimen3\font minus \fontdimen4\font\relax}
\providecommand{\BIBforeignlanguage}[2]{{%
\expandafter\ifx\csname l@#1\endcsname\relax
\typeout{** WARNING: IEEEtran.bst: No hyphenation pattern has been}%
\typeout{** loaded for the language `#1'. Using the pattern for}%
\typeout{** the default language instead.}%
\else
\language=\csname l@#1\endcsname
\fi
#2}}
\providecommand{\BIBdecl}{\relax}
\BIBdecl

\bibitem{ghizzo2025assessing}
E.~Ghizzo \emph{et~al.}, ``Assessing jamming and spoofing impacts on {GNSS} receivers: Automatic gain control ({AGC}),'' \emph{Signal Processing}, vol. 228, p. 109762, 2025.

\bibitem{borio2016impact}
D.~Borio \emph{et~al.}, ``Impact and detection of {GNSS} jammers on consumer grade satellite navigation receivers,'' \emph{Proc. IEEE}, vol. 104, no.~6, pp. 1233--1245, 2016.

\bibitem{spanghero2025gnss}
M.~Spanghero \emph{et~al.}, ``{GNSS} jammer localization and identification with airborne commercial {GNSS} receivers,'' \emph{IEEE Trans. Inf. Forensics Secur.}, 2025.

\bibitem{chen2021carrier}
L.~Chen \emph{et~al.}, ``Carrier phase ranging for indoor positioning with {5G NR} signals,'' \emph{IEEE Internet Things J.}, vol.~9, no.~13, pp. 10\,908--10\,919, 2021.

\bibitem{2019TDOAPSSSSS}
Q.~Liu \emph{et~al.}, ``Simulation and analysis of device positioning in {5G} ultra-dense network,'' in \emph{IWCMC}, 2019, pp. 1529--1533.

\bibitem{ferre2019positioning}
R.~M. Ferre \emph{et~al.}, ``Positioning reference signal design for positioning via {5G},'' in \emph{Finnish URSI Convention on Radio Science}.\hskip 1em plus 0.5em minus 0.4em\relax URSI, 2019.

\bibitem{2017AOATOA}
M.~Koivisto \emph{et~al.}, ``Joint device positioning and clock synchronization in {5G} ultra-dense networks,'' \emph{IEEE Trans. Wireless Commun.}, vol.~16, no.~5, pp. 2866--2881, 2017.

\bibitem{2020HenkAOATDOA}
P.~Gertzell \emph{et~al.}, ``{5G} multi-{BS} positioning with a single-antenna receiver,'' in \emph{IEEE PIMRC}, 2020, pp. 1--5.

\bibitem{2022AOATOA}
H.~Kim \emph{et~al.}, ``Cooperative localization with constraint satisfaction problem in {5G} vehicular networks,'' \emph{Trans. Intell. Transp. Syst.}, vol.~23, no.~4, pp. 3180--3189, 2022.

\bibitem{malmstrom20195g}
M.~Malmstr{\"o}m \emph{et~al.}, ``{5G} positioning-a machine learning approach,'' in \emph{16th Workshop on Positioning, Navigation and Communications (WPNC)}.\hskip 1em plus 0.5em minus 0.4em\relax IEEE, 2019, pp. 1--6.

\bibitem{2020Fingerprinting}
M.~M. Butt \emph{et~al.}, ``{RF} fingerprinting and deep learning assisted ue positioning in {5G},'' in \emph{IEEE VTC-Spring}, 2020, pp. 1--7.

\bibitem{gante2020deep}
J.~Gante \emph{et~al.}, ``Deep learning architectures for accurate millimeter wave positioning in {5G},'' \emph{Neural Processing Letters}, vol.~51, no.~1, pp. 487--514, 2020.

\bibitem{stephan2024angle}
P.~Stephan \emph{et~al.}, ``Angle-delay profile-based and timestamp-aided dissimilarity metrics for channel charting,'' \emph{IEEE Trans. Commun.}, 2024.

\bibitem{fouda}
A.~Fouda \emph{et~al.}, ``Toward cm-level accuracy: Carrier phase positioning for {IIoT} in {5G}-advanced {NR} networks,'' in \emph{IEEE PIMRC}, 2022, pp. 782--787.

\bibitem{wang2023recent}
Y.~Wang \emph{et~al.}, ``Recent progress on {3GPP 5G} positioning,'' in \emph{IEEE VTC-Spring}.\hskip 1em plus 0.5em minus 0.4em\relax IEEE, 2023, pp. 1--6.

\bibitem{Cha2025}
H.-S. Cha \emph{et~al.}, ``{5G NR} positioning enhancements in {3GPP Release-18},'' \emph{IEEE Commun. Stds. Mag.}, vol.~9, no.~1, pp. 22--27, 2025.

\bibitem{10644093}
L.~Italiano \emph{et~al.}, ``A tutorial on {5G} positioning,'' \emph{IEEE Commun. Surveys Tuts.}, pp. 1--1, 2024.

\bibitem{nikonowicz2024}
J.~Nikonowicz \emph{et~al.}, ``Indoor positioning in {5G}-advanced: Challenges and solution toward centimeter-level accuracy with carrier phase enhancements,'' \emph{IEEE Wireless Commun.}, vol.~31, no.~4, pp. 268--275, 2024.

\bibitem{Abuyaghi2025}
M.~Abuyaghi \emph{et~al.}, ``Positioning in {5G} networks: Emerging techniques, use cases, and challenges,'' \emph{IEEE Internet Things J.}, vol.~12, no.~2, pp. 1408--1427, 2025.

\bibitem{Saikko2025}
A.~Saikko \emph{et~al.}, ``High-precision {3D} location and orientation tracking using multi-sensor cellular carrier phase measurements,'' in \emph{33rd European Signal Processing Conference (EUSIPCO)}, 2025, pp. 915--919.

\bibitem{fan2021carrier}
S.~Fan \emph{et~al.}, ``Carrier phase-based synchronization and high-accuracy positioning in {5G} new radio cellular networks,'' \emph{IEEE Trans. Commun.}, vol.~70, no.~1, pp. 564--577, 2021.

\bibitem{kitti}
A.~Geiger \emph{et~al.}, ``Vision meets robotics: The {KITTI} dataset,'' \emph{Int. J. Rob. Res.}, vol.~32, no.~11, p. 1231–1237, Sep. 2013.

\bibitem{hoydis2023sionna}
J.~Hoydis \emph{et~al.}, ``Sionna {RT}: Differentiable ray tracing for radio propagation modeling,'' in \emph{IEEE GC Wkshps}.\hskip 1em plus 0.5em minus 0.4em\relax IEEE, 2023, pp. 317--321.

\bibitem{3GPP_TR38_859}
{3GPP TR 38.859}, ``Study on expanded and improved {NR} positioning,'' June 2024.

\bibitem{Ou2024}
J.~Ou \emph{et~al.}, ``Single-shot carrier phase positioning method with wrapping effect solution in {5G} new radio cellular networks,'' in \emph{IEEE GLOBECOM}, 2024, pp. 313--318.

\bibitem{Deng2024}
Z.~Deng and Z.~Ma, ``A low complexity localization method based on {5G} carrier phase,'' in \emph{9th Intl. Conf. on Intelligent Computing and Signal Processing (ICSP)}, 2024, pp. 872--877.

\bibitem{Wymeersch2023}
H.~Wymeersch \emph{et~al.}, ``Fundamental performance bounds for carrier phase positioning in cellular networks,'' in \emph{IEEE GLOBECOM}, 2023, pp. 7478--7483.

\bibitem{shah2025}
S.~S. Shah \emph{et~al.}, ``Evaluation of {5G} positioning based on uplink {SRS} and downlink {PRS} under {LOS and NLOS} environments,'' \emph{Applied Sciences}, vol.~15, no.~14, 2025.

\bibitem{3gpp_tr_38_857_v17_0_0}
{3GPP TR 38.857}, ``Study on nr positioning enhancements,'' March 2021, release 17.

\bibitem{pan2025ai}
G.~Pan \emph{et~al.}, ``Ai-driven wireless positioning: Fundamentals, standards, state-of-the-art, and challenges,'' \emph{IEEE Communications Surveys \& Tutorials}, 2025.

\bibitem{Yu}
K.~Yu and Y.~J. Guo, ``Statistical nlos identification based on aoa, toa, and signal strength,'' \emph{IEEE Transactions on Vehicular Technology}, vol.~58, no.~1, pp. 274--286, 2009.

\bibitem{si}
M.~Si \emph{et~al.}, ``A lightweight {CIR-Based CNN} with {MLP} for {NLOS/LOS} identification in a {UWB} positioning system,'' \emph{IEEE Commun. Lett.}, vol.~27, no.~5, pp. 1332--1336, 2023.

\bibitem{zhu}
Y.~Zhu \emph{et~al.}, ``A simple efficient lightweight {CNN} method for {LOS/NLOS} identification in wireless communication systems,'' \emph{IEEE Commun. Lett.}, vol.~27, no.~6, pp. 1515--1519, 2023.

\bibitem{wang}
J.~Wang \emph{et~al.}, ``Multi-classification of {UWB} signal propagation channels based on one-dimensional wavelet packet analysis and {CNN},'' \emph{IEEE Trans. Veh. Technol.}, vol.~71, no.~8, pp. 8534--8547, 2022.

\bibitem{Jiang}
C.~Jiang \emph{et~al.}, ``An {UWB} channel impulse response de-noising method for {NLOS/LOS} classification boosting,'' \emph{IEEE Commun. Lett.}, vol.~24, no.~11, pp. 2513--2517, 2020.

\bibitem{vins-mono}
T.~Qin \emph{et~al.}, ``{VINS-Mono}: A robust and versatile monocular visual-inertial state estimator,'' \emph{IEEE Transactions on Robotics}, vol.~34, no.~4, pp. 1004--1020, 2018.

\bibitem{VIO-nonlinear-optimization}
E.~Hong and J.~Lim, ``Visual inertial odometry using coupled nonlinear optimization,'' in \emph{2017 IEEE/RSJ International Conference on Intelligent Robots and Systems (IROS)}, 2017, pp. 6879--6885.

\bibitem{MSCKF-vio}
A.~I. Mourikis and S.~I. Roumeliotis, ``A multi-state constraint {Kalman} filter for vision-aided inertial navigation,'' in \emph{IEEE International Conference on Robotics and Automation}, 2007, pp. 3565--3572.

\bibitem{scaramuzza2019visualinertialodometryaerialrobots}
\BIBentryALTinterwordspacing
D.~Scaramuzza and Z.~Zhang, ``Visual-inertial odometry of aerial robots,'' 2019. [Online]. Available: \url{https://arxiv.org/abs/1906.03289}
\BIBentrySTDinterwordspacing

\bibitem{ml-vio}
\BIBentryALTinterwordspacing
Y.~Pan \emph{et~al.}, ``Adaptive {VIO}: Deep visual-inertial odometry with online continual learning,'' 2024. [Online]. Available: \url{https://arxiv.org/abs/2405.16754}
\BIBentrySTDinterwordspacing

\bibitem{uwb_vio}
\BIBentryALTinterwordspacing
G.~Delama \emph{et~al.}, ``{UVIO}: An {UWB}-aided visual-inertial odometry framework with bias-compensated anchors initialization,'' 2023. [Online]. Available: \url{https://arxiv.org/abs/2308.00513}
\BIBentrySTDinterwordspacing

\bibitem{giordani2020toward}
M.~Giordani \emph{et~al.}, ``Toward {6G} networks: Use cases and technologies,'' \emph{IEEE Commun. Mag.}, vol.~58, no.~3, pp. 55--61, 2020.

\bibitem{wymeersch20226g}
H.~Wymeersch \emph{et~al.}, ``{6G} radio requirements to support integrated communication, localization, and sensing,'' in \emph{EuCNC/6G Summit}.\hskip 1em plus 0.5em minus 0.4em\relax IEEE, 2022, pp. 463--469.

\bibitem{del2017survey}
J.~A. del Peral-Rosado \emph{et~al.}, ``Survey of cellular mobile radio localization methods: From {1G to 5G},'' \emph{IEEE Commun. Surveys Tuts.}, vol.~20, no.~2, pp. 1124--1148, 2017.

\bibitem{3gpp_ts_38_305_v18_3_0}
{3GPP TS 38.305}, ``{{5G}; {NG Radio Access Network} (NG-RAN); Stage 2 functional specification of User Equipment ({UE}) positioning in {NG-RAN}},'' 2024, v18.3.0, Release 18.

\bibitem{3gpp_ts_38_214_v18_4_0}
{3GPP TS 38.214}, ``{5G}; {NR}; physical layer procedures for data,'' Dec. 2024, v18.4.0, Release 18.

\bibitem{3gpp_ts_38_211_v18_4_0}
{3GPP TS 38.211}, ``{5G}; {NR}; physical channels and modulation,'' Dec. 2024, release 18.

\bibitem{3GPP_TR21_918}
{3GPP TR 21.918}, ``Release 18 description; summary of {Rel}-18 work items,'' Mar. 2024.

\bibitem{3GPP_TR38_901}
{3GPP TR 38.901}, ``Study on channel model for frequencies from 0.5 to 100 {GHz},'' Mar. 2022.

\bibitem{Li2021_GNSSRepeater}
X.~Li, ``{GNSS} repeater based differential indoor positioning with multi-epoch measurements,'' \emph{IEEE Trans. Intell. Veh.}, vol.~8, no.~1, pp. 1--813, oct 2021.

\bibitem{3GPP_R1-1901980}
{3GPP Tdoc R1-1901980}, ``Further discussion of {NR} {RAT}-dependent {DL} positioning,'' Mar. 2019.

\bibitem{R1-2306873}
{3GPP Tdoc R1-2306873}, ``Remaining issues on {NR DL} and {UL} carrier phase positioning,'' August 2023.

\bibitem{kabir2021csi}
M.~H. Kabir \emph{et~al.}, ``{CSI-IANet}: An inception attention network for human-human interaction recognition based on {CSI} signal,'' \emph{IEEE Access}, vol.~9, pp. 166\,624--166\,638, 2021.

\bibitem{dfvo-vio}
\BIBentryALTinterwordspacing
H.~Zhan \emph{et~al.}, ``{DF-VO}: What should be learnt for visual odometry?'' 2021. [Online]. Available: \url{https://arxiv.org/abs/2103.00933}
\BIBentrySTDinterwordspacing

\end{thebibliography}








\vfill

\end{document}